\documentclass[aps,reprint,amsmath,amssymb,superscriptaddress,floatfix]{revtex4-1}

\usepackage[colorlinks,citecolor=blue,linkcolor=blue]{hyperref}
\usepackage{graphicx}% Include figure files
\usepackage{dcolumn}% Align table columns on decimal point
\usepackage{bm}% bold math
\usepackage{txfonts}
\usepackage{float}
\usepackage{upgreek}

\newcommand{\fgyr}{\ensuremath{f_\mathrm{gyr}}}
\newcommand{\fext}{\ensuremath{f_\mathrm{ext}}}
\newcommand{\fcr}{\ensuremath{f_\mathrm{cr}}}
\newcommand{\fri}{\ensuremath{f_k}}
\newcommand{\Idc}{\ensuremath{I_\mathrm{DC}}}
\newcommand{\iac}{\ensuremath{i_\mathrm{AC}}}
\newcommand{\ogyr}{\ensuremath{\omega_\mathrm{gyr}}}
\newcommand{\oext}{\ensuremath{\omega_\mathrm{ext}}}

\newcommand{\oi}{\ensuremath{\omega_k}}
\newcommand{\sumi}{\ensuremath{\sum_k}}
\newcommand{\dri}{\ensuremath{\Delta R_k}}
\newcommand{\CNN}{Centre de Nanosciences et de Nanotechnologies, CNRS, Univ. Paris-Sud, Universit\'e Paris-Saclay, 91120 Palaiseau, France}
\newcommand{\UMPhy}{Unit\'e Mixte de Physique, CNRS, Thales, Universit\'e Paris-Sud, Universit\'e Paris-Saclay, 91767 Palaiseau, France}
\newcommand{\IJL}{Institut Jean Lamour, CNRS, Universit\'e de Lorraine, 54011 Nancy, France}

\begin{document}

\title{Modulation and phase-locking in nanocontact vortex oscillators}

\author{J{\'e}r{\'e}my L{\'e}tang}
\affiliation{\CNN}
\author{S{\'e}bastien Petit-Watelot}
\affiliation{\IJL}
\author{Myoung-Woo Yoo}
\author{Thibaut Devolder}
\affiliation{\CNN}
\author{Karim Bouzehouane}
\affiliation{\UMPhy}
\author{Vincent Cros}
\affiliation{\UMPhy}
\author{Joo-Von Kim}
\affiliation{\CNN}

\date{\today}

\begin{abstract}
We have conducted experiments to probe how the dynamics of nanocontact vortex oscillators can be modulated by an external signal. We explore the phase-locking properties in both the commensurate and chaotic regimes, where chaos appears to impede phase-locking while a more standard behavior is seen in the commensurate phase. These different regimes correspond to how the periodicity of the vortex core reversal relates to the frequency of core gyration around the nanocontact; a commensurate phase appears when the reversal rate is an integer fraction of the gyration frequency, while a chaotic state appears when this ratio is irrational. External modulation where the power spectral density exhibits rich features, appears due to the modulation between the external source frequency, gyration frequency, and core-reversal frequency. We explain these features with first- or second-order modulation between the three frequencies. Phase-locking is also visible between the external source frequency and internal vortex modes (gyration and core reversal modes).
\end{abstract}

\maketitle

%%
%	Section: Introduction
%%
\section{Introduction} 
Spin-torque nano-oscillators \cite{slonczewski_current-driven_1996, berger_emission_1996, kiselev_microwave_2003, slavin_nonlinear_2009, bonetti_direct_2015} (STNO) have strong potential for applications such as rf communications, microwave generation \cite{kreissig_vortex_2017}, field sensing \cite{tulapurkar_spin-torque_2005}, and neuro-inspired computing~\cite{torrejon_neuromorphic_2017, romera_vowel_2018, tsunegi_physical_2019}. An important aspect involves phase-locking \cite{rippard_injection_2005, puliafito_synchronization_2014, lebrun_understanding_2015,  gopal_phase_2019}  and modulation \cite{pufall_frequency_2005, consolo_combined_2010, muduli_nonlinear_2010, pogoryelov_spin-torque_2011, martin_tunability_2013} with external signals, which have been studied extensively in vortex-based systems~\cite{dussaux_phase_2011, hamadeh_perfect_2014}. However, the role of vortex core reversal~\cite{van_waeyenberge_magnetic_2006} in this context has remained largely unexplored. Indeed, in nanocontact-based systems, core reversal can give rise to more complex states such as rich modulation patterns but also a chaotic dynamic~\cite{petit-watelot_commensurability_2012, pylypovskyi_regular_2013, williame_chaotic_2019, bondarenko_chaotic_2019, taniguchi_synchronized_2019, matsumoto_chaos_2019}. Because of the sensitivity to initial conditions, chaos is potentially useful for information processing as a large number of patterns can be generated rapidly~\cite{sciamanna_physics_2015}, and therefore be used as a random number generator, in symbolic dynamics or even neuromorphic computing.

The difference between nanopillar~\cite{martin_parametric_2011} and nanocontact~\cite{manfrini_frequency_2011} systems in term of modulation and phase-locking originates mainly in the geometry. For vortex-based oscillators, the geometry is important because different spin-torque components are at play. In the nanopillar geometry, the primary contribution to dynamics arises from spin-transfer torques of the Slonczewski form due to the current flowing perpendicular to the film plane (CPP)~\cite{slonczewski_current-driven_1996, berger_emission_1996}, where the out-of-plane component of the spin torque determines which sense of gyration is possible relative to the core polarity, $p$~\cite{ivanov_excitation_2007}. In other words, self-gyration of the vortex core is only possible for certain combinations of the current density $J$, the polarity $p$, and the spin polarization direction $p_z$. The condition for oscillations is $J p p_z > 0$. This is in stark contrast to the nanocontact system, where the primary driving torques in a steady state are of the Zhang-Li form~\cite{zhang_roles_2004} due to currents flowing in the film plane (CIP). As such, there is no condition on the core polarity for self-sustained gyration, which allows for phenomena such as periodic core reversal to occur~\cite{petit-watelot_commensurability_2012}.

In this paper, we present an experimental and theoretical study in which we investigated how modulation and phase-locking due to the injection of an external current affect the vortex dynamics in a nanocontact oscillator. Particular focus is given to the core-reversal regime, where periodic core reversal occurs in addition to the usual vortex gyration around the nanocontact~\cite{petit-watelot_commensurability_2012}. A notable feature is the existence of both commensurate states, where the ratio between the core-reversal and gyration frequencies is an integer fraction, and incommensurate or chaotic states, where this ratio is irrational. We find that external modulation affects these two periodic processes differently, which offers insight into how chaos may be induced and controlled in such oscillators.

This paper is organized as follows. In Sec. II, we describe the materials and sample fabrication, the experimental setup for the electrical measurements, and the simulation methods employed. In Sec. III, an overview is given of the three oscillator regimes studied. In Sec. IV, the response of the nanocontact vortex oscillator to alternating currents in the different regimes is presented. In Sec. V, we describe the effects of current modulation on the periodic core reversal. A discussion and concluding remarks are given in Sec. VI.

%%
% Section
%%
\section{Experimental setup and simulation methods} 

\subsection{Materials and sample fabrication}
\begin{figure}
%VPBC8
\centering\includegraphics[width=7.5cm]{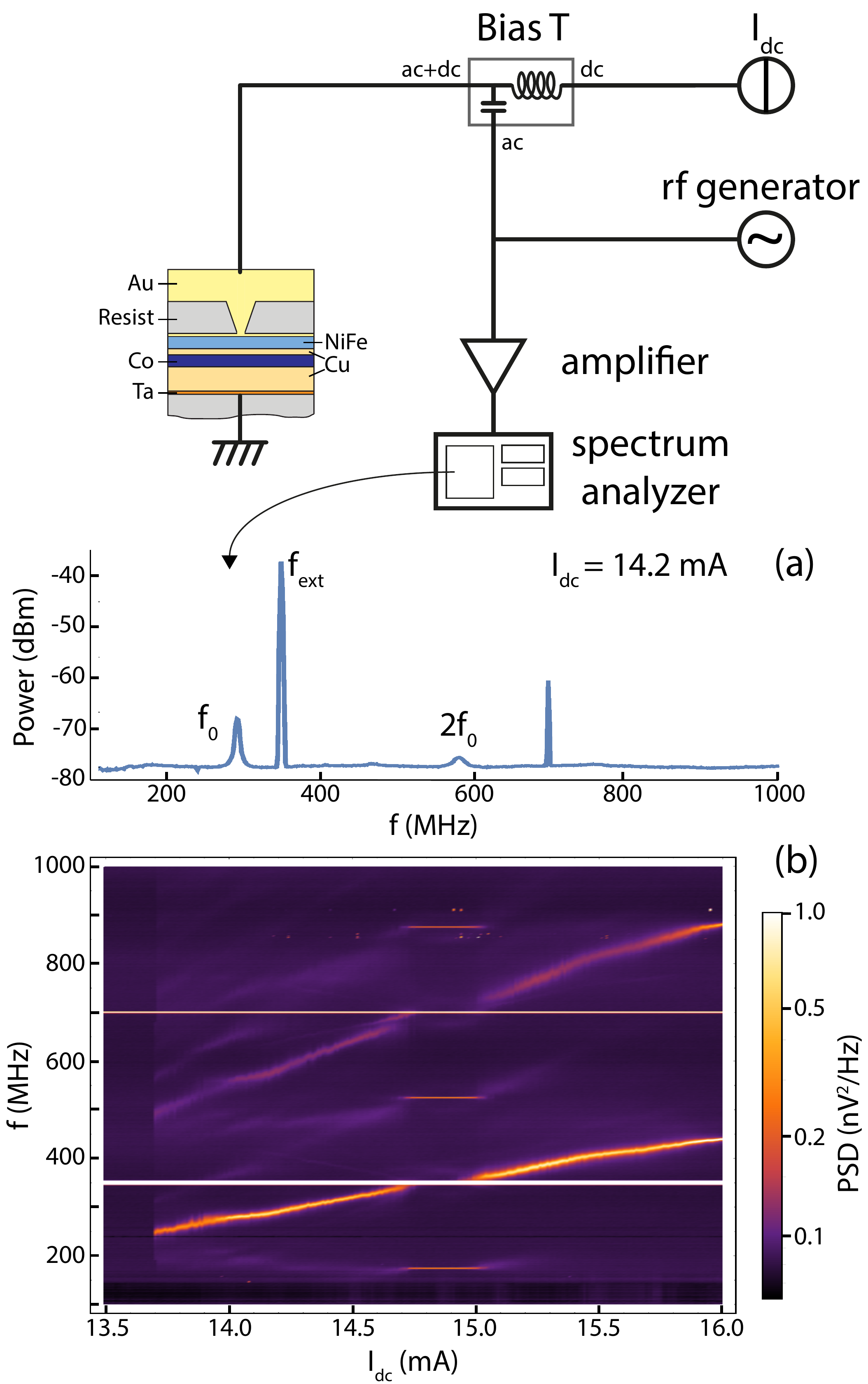}
\caption{Experimental set up and device geometry. The current flows from the top electrode into the multilayer stack, until the other electrode. (a) Spectrum as read on the spectrum analyzer for a DC current of 14.2 mA. After treatment, spectra are aggregated to give (b) a PSD map.}
\label{fig:setup}
\end{figure}

The oscillator comprises a metallic nanocontact adjacent to a pseudo spin valve, as illustrated in Fig.~\ref{fig:setup}. 
The multilayer is deposited using sputtering and has the composition Ta(5)/Cu(40)/Co(20)/Cu(10)/NiFe(20)/Au(5), where the figures in parentheses denote the film thickness in nm. An insulating resist layer is then deposited on top of the Au cap layer, through which a nanocontact is formed using a nano-indentation technique involving the conductive tip of an atomic force microscope~\cite{bouzehouane_nanolithography_2003}. The nanocontact has the shape of a truncated pyramid, with a lateral size of approximately 20~nm in contact with the spin valve, 40~nm thick. Electrical measurements are made possible via gold electrodes to the nanocontact. Further details on the fabrication technique can be found in previous work~\cite{ruotolo_phase-locking_2009, petit-watelot_commensurability_2012}.

The NiFe layer is the free layer in which the vortex dynamics takes place, while the Co layer is the reference magnetic layer allowing the giant magnetoresistance signal. The magnetic properties of the films before patterning were determined with vector network analyzer ferromagnetic resonance before patterning. The NiFe layer is found to have a coercivity of 1~mT, a saturation magnetization of $1.053 \pm 0.003$~T, a spectroscopic splitting factor ($g$-factor) of $2.111 \pm 0.003$, and a Gilbert damping constant of $(7 \pm 1).10^{-3}$. The Co layer is also relatively soft with a coercivity of 2~mT, a saturation magnetization of $1.768 \pm 0.011$~T, a $g$-factor of $2.133 \pm 0.009$, and a Gilbert constant of $(10 \pm 1).10^{-3}$ without any inhomogeneous broadening of the ferromagnetic resonance line. These parameters correspond to a polycrystalline cobalt film with an fcc structure.

\subsection{Electrical measurements}
%Explication rapide du principe de la mesure
The electrical measurements of the nanocontact device have been performed at 77~K. The main contribution to the signal arises from the gyrotropic dynamics of the magnetic vortex that is induced by the applied DC currents. The gyration leads to a resistance variation that translates into voltage oscillations in the sub-GHz range. These oscillations were generally measured with a spectrum analyzer in the range of 100 to 1000~MHz. The spectrum analyzer was mainly used with a resolution bandwidth of 50~kHz, a video bandwidth of 5~kHz. The signal is amplified with a 50~dB broadband amplifier before being fed into the spectrum analyzer. In addition to the DC current, we also apply a radiofrequency current with a synthesizer as an additional modulation. The circuit is illustrated in Fig. \ref{fig:setup}. For the majority of the measurements here, we fix either the DC current or the frequency of the modulation signal, with the other parameter being varied.

An example is given in Fig.~\ref{fig:setup}(b), where the power spectral density (PSD) is shown as a color map as a function of the DC current $\Idc$ at a fixed value of external modulation frequency [white line in Fig.~\ref{fig:setup}(b)]. The PSD at each current is represented using a color code, which allows features in the power spectrum to be followed as $\Idc$ is varied. For the sake of brevity, such plots are referred to as maps in this paper.

\subsection{Electrical and micromagnetics simulations}
It has previously been shown that an accurate description of the electrical current and associated {\O}rsted-Amp{\`e}re field profiles in the nanocontact geometry is necessary to provide a good quantitative agreement with experimental observations~\cite{jaromirska_influence_2011, petit-watelot_understanding_2012, petit-watelot_electrical_2012}. To this end, we have employed the finite-element code \textsc{Comsol} to compute the current and {\O}rsted-Amp{\`e}re field profiles in the nanocontact devices studied using the method described in Ref.~\cite{petit-watelot_understanding_2012}. By assuming cylindrical symmetry, we model the multilayer cross section as a 2~$\upmu$m $\times$ 100~nm rectangle with the nanocontact at one end. The full multilayer stack is simulated with the bulk values of the conductivity used for each material. The nanocontact itself is taken to be a right trapezium whose 13.5~nm smaller parallel side is in contact with the multilayer stack~\cite{petit-watelot_electrical_2012}. Temperature and electrical wave propagation effects have been neglected in this calculation. The simulations give the dependence of the perpendicular-to-plane and in-plane current densities as a function of radial distance from the nanocontact and film thickness in the ferromagnetic free layer. Since we neglect the thickness dependence when considering the magnetization dynamics, the current and {\O}rsted-Amp{\`e}re field profiles are averaged over the film thickness of the free layer. With \textsc{Comsol}, we have calculated that the {\O}rsted-Amp{\`e}re field is around 800~A/m for a DC current of 1 mA, which corresponds to an increase of 1~mT for every 1~mA. This field is in-plane but arises from both CIP and CPP components\cite{petit-watelot_electrical_2012}.

The magnetization dynamics is studied with the micromagnetics code \textsc{Mumax3}~\cite{vansteenkiste_design_2014}, which performs a numerical time integration of the Landau-Lifshitz-Gilbert equation with spin transfer torques~\cite{landau_theory_1935, gilbert_lagrangian_1955, gilbert_phenomenological_2004},
\begin{equation}
\frac{d \mathbf{m}}{dt} = -\gamma_0 \mathbf{m} \times \mathbf{H}_\mathrm{eff} + \alpha \mathbf{m} \times \frac{d \mathbf{m}}{dt} + \mathbf{\Gamma}_\mathrm{ST}.
\label{eq:LLG}
\end{equation}
$\gamma_0$ is the gyromagnetic ratio, $\mathbf{m}(\mathbf{r},t)$ is a unit vector representing the magnetization field, $\mathbf{H}_\mathrm{eff}$ is the effective magnetic field, $\alpha$ is the Gilbert damping constant, and $\mathbf{\Gamma}_\mathrm{ST}$ represents nonconservative spin transfer torques. The effective field is given by the variational derivative of the total magnetic energy density $U$ with respect to the magnetization unit vector, $\mathbf{H}_\mathrm{eff} = - (1/\mu_0 M_s) \delta U/\delta \mathbf{m}$, and comprises contributions from the exchange, dipole-dipole, and the Zeeman interactions, where the latter includes contributions from the static external applied field and the {\O}rsted-Amp{\`e}re field generated by the current flow through the nanocontact.

The simulation geometry comprises a $1280 \times 1280 \times 20$ nm system that is discretized using $512 \times 512 \times 1$ finite difference cells. We use micromagnetic parameters suitable for Permalloy; we take the saturation magnetization to be $M_{s} = 800$ kA/m, the exchange stiffness $A_{ex} = 10$ pJ/m, and the Gilbert damping constant $\alpha = 0.013$ (a standard value for NiFe). For the spin-transfer torques, the dominant contribution comes from the current in-plane terms, so we use
\begin{equation}
\mathbf{\Gamma}_\mathrm{ST} = -\left[ \mathbf{u}(\mathbf{r}) \cdot \nabla \right] \mathbf{m},
\end{equation}
where $\mathbf{u} = \mathbf{J}(\mathbf{r}) P \mu_B/ (e M_s)$ represents an effective spin drift velocity, where $\mathbf{J}$ is the in-plane current density, $\mu_B$ is the Bohr magneton, $e$ is the electron charge, with the spin polarization taken to be $P = 0.5$. We have verified that the nonadiabatic and Slonczewski terms are negligible in nanocontact vortex dynamics, so no further considerations to these terms will be given here. The spatial profiles for $\mathbf{J}(\mathbf{r})$ and the {\O}rsted-Amp{\`e}re field it generates, $\mathbf{H}_\mathrm{Oe}(\mathbf{r})$, determined using the \textsc{Comsol} simulations described above, are used as inputs into the micromagnetics simulations.

The initial magnetic state in the free layer is obtained by mimicking the experimental procedure, which is described in Ref. \cite{petit-watelot_commensurability_2012} and later in Sec.~\ref{part3}. In this procedure, once the transient dynamics has died out, we obtain a self-gyrating vortex around the nanocontact, with a remnant antivortex structure pinned to one edge of the simulation box~\cite{petit-watelot_commensurability_2012}. This serves as the initial condition for subsequent simulations. For calculations that involve sweeping the applied current, we use the final state of the simulation at a given current value as the initial state for the subsequent value.

The current dependence of the PSD of vortex oscillations is computed as follows. For each value of the applied current, we conduct the simulation over an interval of 100~ns, from which we extract the spatially averaged $m_x$ component, which is representative of the giant magnetoresistance signal, and the core polarity, which is a measure of the core polarity, $p$. Since an adaptative time step is used in the numerical time integration, this data is resampled using cubic interpolation to recreate a time series with equal time steps. Fast Fourier transforms are then applied to this time series data, from which we compute the PSD. In what follows, the PSD under modulation is studied either as a function of DC current, where current steps of 0.1 mA are used, or as a function of the modulation frequency, where frequency steps of 15~MHz are used.

\section{Overview of oscillator regimes\label{part3}}

Experimentally, the ground state of the magnetic free layer is the uniformly-magnetized state. As such, a nucleation procedure is required before measurements to generate the vortex state for self-oscillations. To achieve this, a 10 mT in-plane magnetic field is applied to saturate the magnetization along one direction. A large DC current is then applied (around 16 or 17 mA), which generates a strong {\O}rsted-Amp{\`e}re field around the nanocontact with a circulating profile that favors one vortex chirality. The applied field is then swept quasi-statically to $-10$ mT, during which the magnetization in the free layer reverses through domain wall nucleation and propagation. As the domain wall sweeps through the nanocontact region, a vortex is nucleated~\cite{devolder_vortex_2011}, which results in the well-defined features in the power spectrum [Fig.~\ref{fig:setup}(b)]. After nucleation, we change the applied field to tune the oscillator regime.

This nucleation procedure strongly depends on the initial conditions, so the ease with which nucleation occurs can fluctuate between experiments. To preserve the overall topology of the magnetization state, it is conjectured that the vortex nucleation is always accompanied by the nucleation of an antivortex~\cite{devolder_vortex_2011, nakatani_nucleation_2008}; while the former is attracted to the Zeeman potential associated with the {\O}rsted-Amp{\`e}re field, the latter is repelled by this potential~\cite{otxoa_dynamical_2015}. The presence of an antivortex is supported by the observation of a large number of harmonics~\cite{petit-watelot_commensurability_2012} in the experimental power spectrum at low currents and by simulations. The need for such nucleation processes means that the measured power spectra can exhibit small quantitative differences between successive nucleation events~\cite{devolder_chaos_2019}, however, the power spectra remains unchanged between measurements after a given nucleation event. The measurements are typically done by decreasing the DC current from nucleation current, down to a critical value, where the vortex can be annihilated. If the DC current is kept above this critical value (around 10 mA), vortex annihilation is unlikely and measurements can be performed for both increasing and decreasing current sweeps. Vortex annihilation is certain under 4 mA.

\begin{figure}
\centering\includegraphics[width=8cm]{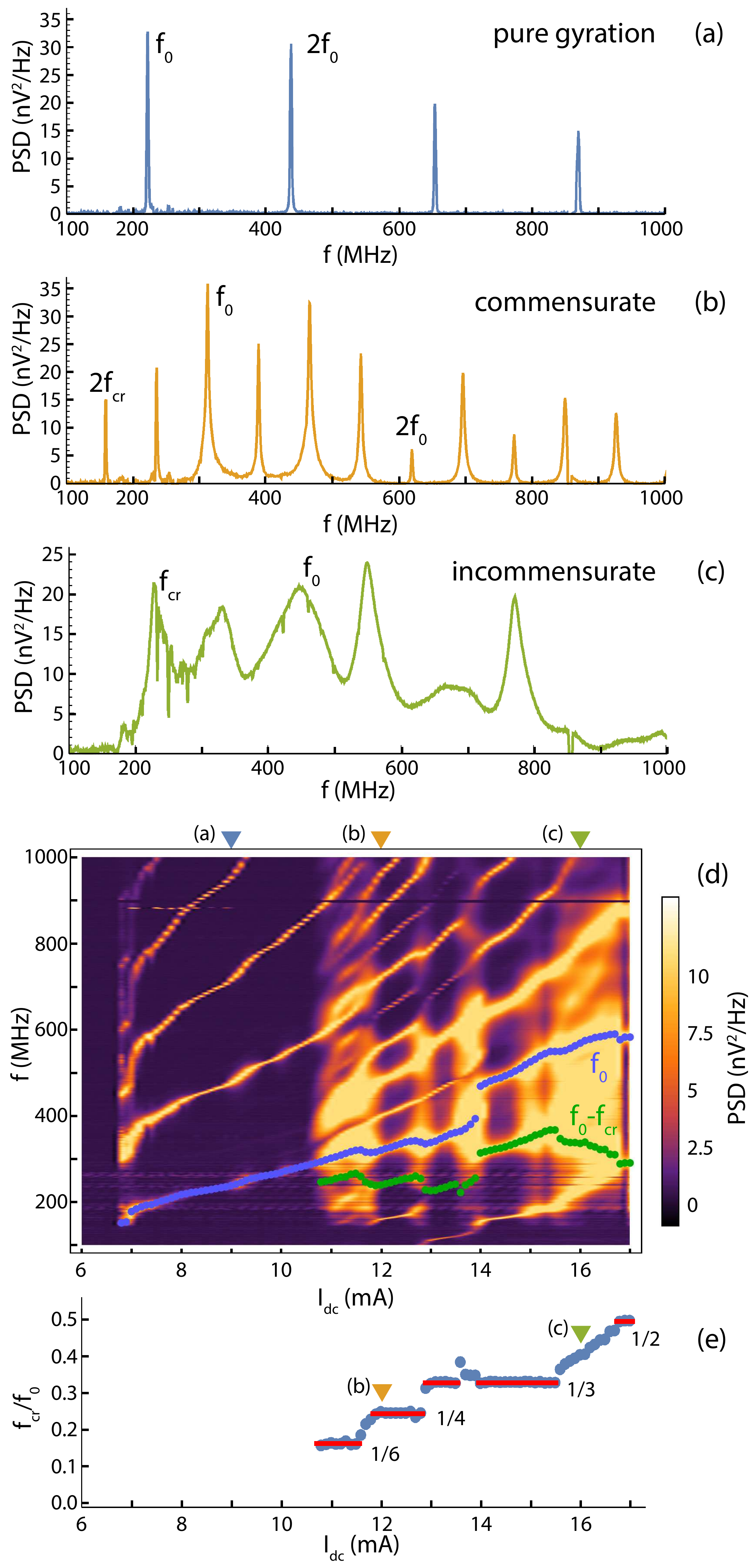}
\caption{Spectra of PSD (in nV$^2$/Hz) vs frequency (in MHz), at 8.5 mA (steady gyration regime) (a), at 12 mA (modulated regime) (b) and at 16 mA (chaotic regime) (c). Aggregated spectra measurements with DC current varying give the PSD map (d). Red triangles give the above spectra correspondence. Upper (blue) dots correspond to gyration frequency. Lower (green) dots correspond to $f_0 - f_{cr}$. (e) gives ratio between $f_{cr}$ and $f_0$ versus DC current.}
\label{fig:spectrajmap}
\end{figure}

An example of the experimental power spectra is shown in Fig.~\ref{fig:spectrajmap}.
The current dependence of the PSD is presented in Fig.~\ref{fig:spectrajmap}(d). Peaks in the sub-GHz range appear above a threshold around 7 mA, where the power is concentrated in the fundamental frequency which is indicated by the blue line. This first oscillation mode corresponds to vortex gyration around the nanocontact. The trajectory of the vortex is conjectured to be noncircular, since a number of higher order harmonics are clearly visible in the power spectrum. An example of the PSD in this regime is shown in Fig.~\ref{fig:spectrajmap}(a),  where these harmonics can be clearly seen. This steady-state gyration regime extends from 7 to 11 mA, though this interval may vary between nucleation events (e.g., from 5 to 13 mA in a different experiment not shown here).

As the current is increased, a second threshold is reached above which periodic core reversal takes place~\cite{petit-watelot_commensurability_2012}. This corresponds to the appearance of additional sidebands in the PSD, which can be observed in Fig.~\ref{fig:spectrajmap}(b) for an applied current of 12 mA. In this figure, the fundamental frequency is labeled by $f_0$ and the core-reversal frequency is labeled by $f_\mathrm{cr}$ (note that $\fcr$ is below the measurement range for a few current values). This example represents a \emph{commensurate} state because the ratio between the core-reversal and gyration frequencies is a rational fraction, as shown in Fig.~\ref{fig:spectrajmap}(e). These ratios vary as the current is increased and the presence of plateaus is indicative of a self-phase-locked state, whereby an integer multiple of the core reversal frequency is locked to the gyration frequency. Physically, this means that core reversal occurs after integer revolutions around the nanocontact.

In between the plateaus, we can observe instances in which the ratio $f_\mathrm{cr}/f_0$ is irrational. A clear example can be seen between the $1/3$ and $1/2$ plateaus in Fig.~\ref{fig:spectrajmap}(e), where this ratio appears to vary linearly with current. An example of the PSD is shown in Fig.~\ref{fig:spectrajmap}(c) at a current of 16 mA. In contrast to the commensurate state, the PSD in this regime is characterized by broad peaks with no obvious relationship between $f_\mathrm{cr}$ and $f_0$. This regime is termed the \emph{incommensurate} state and corresponds to temporal chaos; while core reversal occurs after an integer number of revolutions around the nanocontact, this number itself is characterized by a chaotic sequence~\cite{petit-watelot_commensurability_2012, devolder_chaos_2019}. In other words, the dynamics in this regime is characterized by vortex gyration that is modulated by chaotic vortex core reversal. However, this behavior contrasts with other results~\cite{kuepferling_two_2010, wang_multiple-mode_2011, keatley_direct_2016, keatley_imaging_2017} where two modes coexist without chaos. This is due to two weakly coupled parameters, such as two layers~\cite{kuepferling_two_2010, wang_multiple-mode_2011} or two nanocontacts~\cite{keatley_direct_2016, keatley_imaging_2017}.

\begin{figure}
\centering\includegraphics[width=8.5cm]{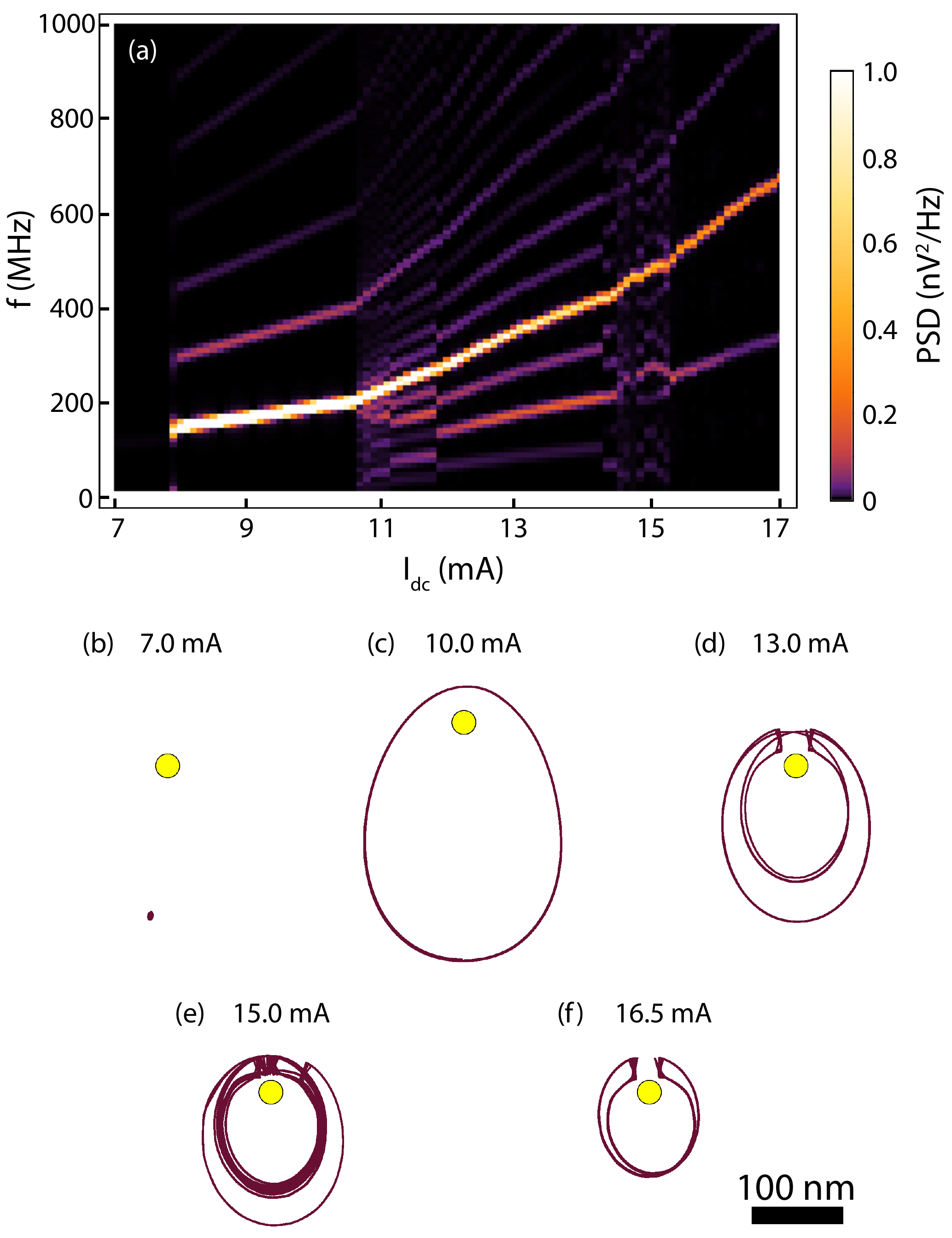}
\caption{Simulated power spectra of vortex gyration, with (a) current dependence of the power spectral density and (b)-(f) show different trajectories: (b) Static vortex core, below the threshold current; (c) Steady-state gyration; (d) Commensurate 1/4 regime; (e) Chaotic regime; (f) Commensurate 1/2 regime. In (b)-(f), an antivortex is situated on same side than the static vortex, much further.}
\label{fig:simutraj}
\end{figure}

The main features of the experimental spectra are reproduced in the micromagnetics simulations. The simulated current dependence of the PSD is presented in Fig.~\ref{fig:simutraj}(a).
We observe a finite lower threshold current for oscillations at 7.9 mA. Below this threshold, the vortex core is immobile and localized at a distance of around 160 nm from the center of the nanocontact, as shown in Fig.~\ref{fig:simutraj}(a). This position results from a competition between the attractive central potential of the Zeeman interaction associated with the {\O}rsted-Amp{\`e}re field, and the attractive interaction between the vortex and the antivortex, where the latter is pinned at the edge of the simulation box~\cite{petit-watelot_commensurability_2012}. The potential asymmetry is also due to a small contribution from CIP currents as discussed in Ref.~\cite{petit-watelot_electrical_2012}. Once this lower threshold is overcome, we observe a steady-state gyration of the vortex around the nanocontact [Fig.~\ref{fig:simutraj}(c)], where the trajectory represents a limit cycle with an egglike form that results from the balance between the asymmetric potential landscape, as discussed above, with the radial symmetry of the spin torques due to the in-plane currents flowing from the nanocontact~\cite{petit-watelot_electrical_2012}. The absence of a radial symmetry of this trajectory gives rises to the rich harmonic content of the power spectra [Fig.~\ref{fig:spectrajmap}(a)].

Core reversal does not occur at arbitrary points along the trajectory, but takes place close to the nanocontact, where the vortex velocity is higher, as shown in Figs.~\ref{fig:simutraj}(d)-\ref{fig:simutraj}(f). The core reversal process involves the strong deformation of the vortex core, where a `dip' in the $m_z$ component is generated in the direction opposite to the core polarity~\cite{yamada_electrical_2007, guslienko_dynamic_2008}. Once a critical deformation is reached, the dip transforms into the nucleation of a vortex-antivortex pair with an opposite polarity, and the original vortex annihilates with the antivortex~\cite{van_waeyenberge_magnetic_2006}, leading to a burst of spin waves~\cite{hertel_exchange_2006}. Because the core reversal process is actually mediated by the annihilation and nucleation of a vortex with an opposite polarity, a discontinuity appears in the core position and is represented by the sharp near-vertical lines in Figs.~\ref{fig:simutraj}(d)-\ref{fig:simutraj}(f) above the nanocontact. The periodic core reversal is analogous to relaxation oscillations; after a reversal, the core spirals outward from the nanocontact center, gaining in energy, until the critical deformation is reached and energy is released at the subsequent reversal~\cite{petit-watelot_commensurability_2012}. We also observe that the trajectories shrink as the current is increased, which results in $f_0$ increasing faster than a linear function in the current, as observed experimentally [Fig.~\ref{fig:spectrajmap}(d)] and in simulation [Fig.~\ref{fig:simutraj}(a)]. We note also that reversals in the core polarity result in a change in the sense of the gyration around the nanocontact (i.e., clockwise or counterclockwise). This leads to the modulation sidebands seen in the power spectra.

For the commensurate states, the trajectories have a clear overlap where the core reversal events occur at near-identical positions, as can be seen for the 1/4 and 1/2 states in Figs.~\ref{fig:simutraj}(d) and \ref{fig:simutraj}(f), respectively. In the chaotic regime, on the other hand, the point at which the core reverses can vary greatly between revolutions around the nanocontact [Fig.~\ref{fig:simutraj}(e)]. This results in a set of trajectories that cover a greater area around the nanocontact, which translates into the broad spectral peaks as seen in Fig.~\ref{fig:spectrajmap}(c). Because of the large qualitative differences in the trajectories between the steady-state, commensurate, and chaotic regimes, we can anticipate that external forcing with AC currents will have different effects on the core dynamics.

%%
% Section
%%
\section{Modulation due to alternating currents}

In this section, we describe the effects of external forcing due to AC currents on the different oscillatory regimes of the nanocontact vortex oscillator. AC currents lead to periodic modulations in the Zeeman potential, associated with the {\O}rsted-Amp{\`e}re field, and to periodic modulations in the spin torques exerted on the vortex core. To see how these terms might influence the core dynamics, consider the Thiele equation, which provides a good description of the gyration far below the threshold for core reversal~\cite{thiele_steady-state_1973, huber_dynamics_1982, Kim:2012du}, 
\begin{equation}
\mathbf{G} \times \left( \dot{\mathbf{X}}_0 - \mathbf{u}(\mathbf{X}_0,I) \right)  + \mathcal{D} \cdot \dot{\mathbf{X}}_0 = -\frac{\partial U(I)}{\partial \mathbf{X}_0}.
\label{eq:Thiele}
\end{equation}
$\mathbf{X}_0$ is the position of the vortex core in the film plane, $\dot{\mathbf{X}}_0 \equiv \partial_t \mathbf{X}_0$, $\mathbf{G}$ is the gyrovector, $\mathcal{D}$ is the Gilbert dissipation tensor, $\mathbf{u}$ is the effective spin drift velocity that measures the strength of the spin torques, $U$ is the total energy of the vortex system, and $I$ is the applied current. This equation of motion can be derived from the Landau-Lifshitz equation [Eq.~(\ref{eq:LLG})] by assuming a rigid core for the vortex. As such, it captures the gyrotropic dynamics but it cannot account for vortex core reversal.

\begin{figure}
\centering\includegraphics[width=8cm]{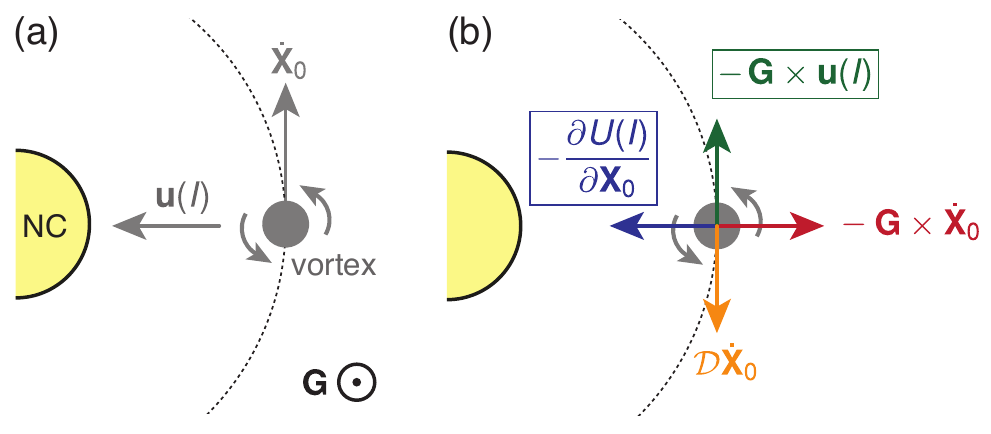}
\caption{Schematic representation of circular vortex gyration around the nanocontact. (a) Vortex gyration in a counterclockwise direction with velocity $\dot{\mathbf{X}}_0$. $u$ indicates the direction of the effect spin drift velocity. (b) Force diagram for the four terms in the Thiele equation [Eq.~(\ref{eq:Thiele})]. The current-dependent terms are outlined by a box.}
\label{fig:force_NC}
\end{figure}

There are two current-dependent terms, the spin current $\mathbf{u}$ and the potential energy density $U$; modulations in the current, $I = I_\mathrm{dc} + i_\mathrm{ac}$, will therefore result in modulations in these two terms.  To see how these enter the dynamics, consider the circular motion around the nanocontact as a result of a pure central potential, $U(\| \mathbf{X}_0  \|)$, i.e., we neglect contributions from exchange interactions with the antivortex. We will also assume a counterclockwise gyration (when viewed from above, $+z$), which corresponds to a gyrovector $G = 2\pi M_s d p /\gamma $ where $d$ is the film thickness and $p$ is the core polarization. A schematic of this motion is given in Fig.~\ref{fig:force_NC}(a).
A pictorial representation of the four force terms in Eq.~($\ref{eq:Thiele}$) is given in Fig.~\ref{fig:force_NC}(b). This figure gives a clear interpretation of how the four forces counterbalance each other. The restoring force due to the Zeeman potential is directed radially inwards, which favors the vortex core centered on the nanocontact, while the gyrotropic term is directed radially outward. The equilibrium orbit is therefore determined by a balance not only between these two forces, but from all forces as two of them share a common term, $\dot{\mathbf{X}}_0$. Modulations in the strength of the Zeeman potential, due to the AC current, amounts to a modulation of the radial force and therefore acts to modulate the gyrovector, therefore the radius of the vortex gyration. Moreover, since the potential also determines the gyration frequency~\cite{ivanov_excitation_2007, Kim:2012du}, this modulation is akin to a parametric excitation. Let us now discuss the two other forces: damping and spin torque. Both act tangentially to the circular orbit, where the damping acts like friction in the direction opposite to the motion, while the adiabatic spin torque term acts in the direction of the motion as a velocity ``boost''. Compensation between these two is required for the vortex to maintain steady-state gyration around the nanocontact. Modulations in the current lead to a modulation in the adiabatic torque, which acts to modulate the ``boost'' of the vortex core along its trajectory. This is akin to a phase modulation of the vortex oscillator. In the following, we will discuss how these two contributions affect the vortex dynamics in the three regimes.

\subsection{Forcing in the steady gyration regime}
We first examine the effects of current modulation in the low current regime in which no core reversal is present. The experimental power spectra are presented in Fig.~\ref{fig:allforc}(a), which correspond to the following operating conditions:
\begin{figure*}
% fitvp64			| VPCE23			| VPCE17
% hzm1p1		| fabuleux			| hz4v
% 117synchro	| 115synchro	| 119synchro
\centering\includegraphics[width=18cm]{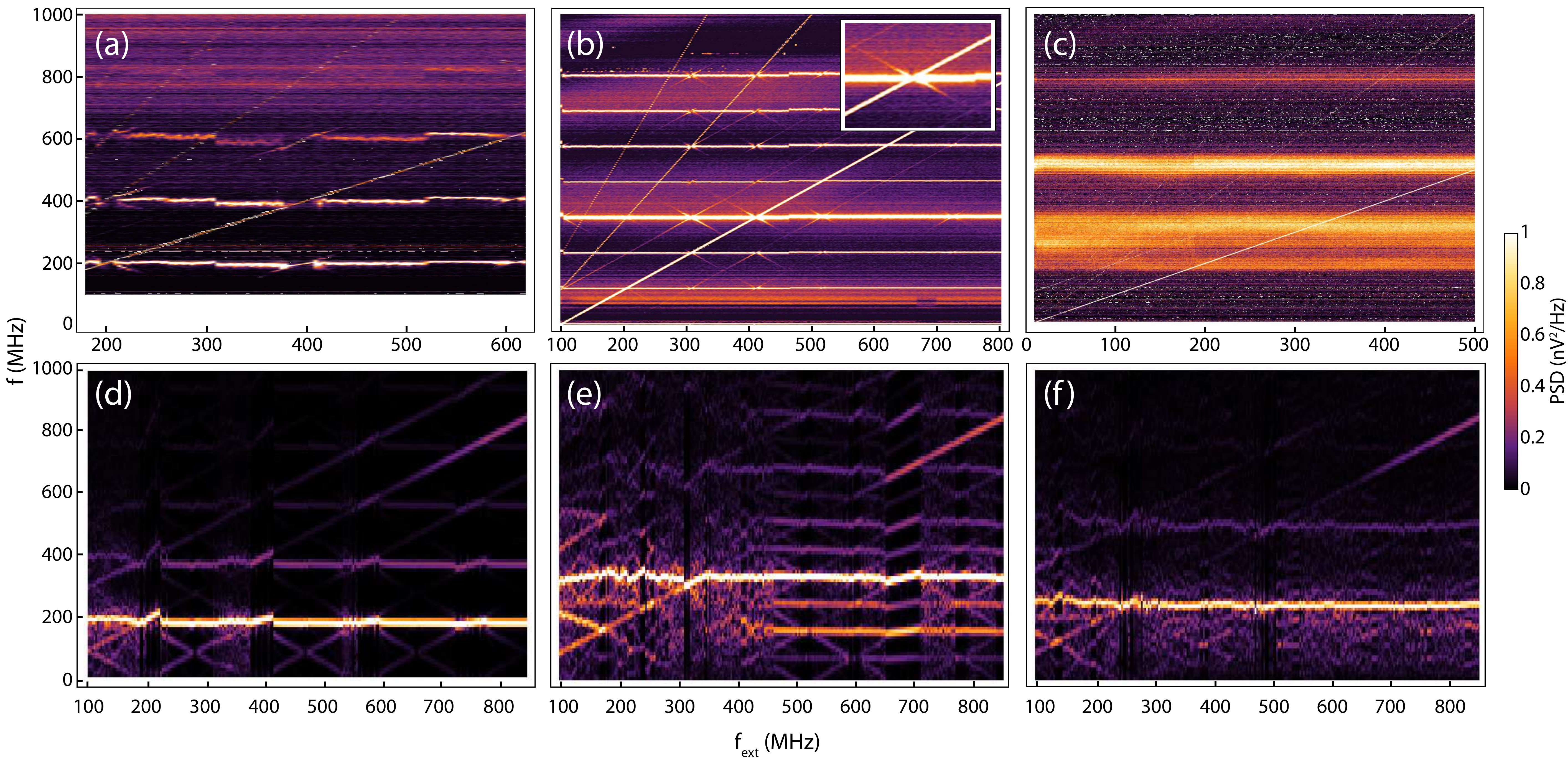}
\caption{Map of the power spectral density as a function of modulation frequency. Parts (a)-(c) correspond to experimental measurements, while (d)-(f) correspond to results of micromagnetics simulations. Left [(a),(d)], central [(b),(e)], and right [(c),(f)] columns correspond to the pure gyration regime, core modulated regime, and chaotic regime, respectively. The inset in (b) is a zoom on the region between 350 and 450 MHz on both scales; 1:1 phase-locking and modulation sidebands are visible. The applied current $\Idc$, in-plane field $H$, and perpendicular-to-plane field $H_{\perp}$ for each part are as follows: (a) $\Idc = 12.8$ mA, $\mu_0 H = 294$ $\upmu$T; (b) $\Idc = 15$ mA, $\mu_0 H = 68$ $\upmu$T; (c) $\Idc = 16.7$ mA, $\mu_0 H_{\perp} = 90.4$ mT; (d) $\Idc = 10$ mA, $\mu_0 H = 2$ mT; (e) $\Idc = 13$ mA, $\mu_0 H = 2$ mT; (f) $\Idc = 11.5$ mA, $\mu_0 H = 2$ mT.}
\label{fig:allforc}
\end{figure*}
a DC current of 12.8 mA in Fig.~\ref{fig:allforc}(a) resulting in a gyration frequency of 200 MHz. An AC modulation of $i_\mathrm{ac} = 0.3$~mA (peak to peak) is applied whose frequency $f_\mathrm{ext}$ is swept between 180 and 620 MHz, which is clearly present in the experimental spectra as a narrow line with unity slope, $f = \fext$. %The ac current amplitude is limited by the ground noise of the synthesizer, which create the external frequency. 
We note that harmonics in the external forcing can also be seen (i.e., fainter lines with $f = 2\fext$ and $f = 3\fext$). We attribute this to nonlinearities in the gain of the amplifiers we used, which means the sample does not receive those harmonics.

Phase locking to the external signal can be seen in Fig.~\ref{fig:allforc}(a) as $\fext$ crosses $\fgyr$ at around 200 MHz, which is evidenced by a vacant horizontal segment in the power spectrum at which the oscillator frequency is entrained by the external signal. Because of the elliptical trajectory of the vortex core around the nanocontact, this entrainment also manifests itself in the higher harmonics, notably at $2 \fgyr$. We also note that fractional synchronization is seen in Fig.~\ref{fig:allforc}(a), whereby phase locking occurs at integer multiples or integer fractions of the gyration frequency. Phase-locking is perceptible at $\fext = \fgyr$, $\fext = 2\fgyr$ and $\fext = 3 \fgyr$. %The vortex frequency $\fgyr$ exhibits some instabilities. 
Overall, this behavior is similar to the phase-locking phenomenon observed in nanopillar vortex oscillators~\cite{dussaux_phase_2011, hamadeh_perfect_2014}, though it seems we do not have fractional phase-locking in this regime as in nanopillar oscillators~\cite{urazhdin_fractional_2010}.

To understand the modulation process in this pure gyration case, we can use a simple model of $r(t)$ the resistance of the system and $i(t)$ the current flowing through the system
\begin{equation}
\begin{split}
r(t) &= R_0 + \Delta R \cos (\ogyr t), \\
i(t) &= \Idc + \iac \cos (\oext t), 
\end{split}
\end{equation}
where $R_0$ is the mean resistance of the device, $\Delta R$ is the resistance variation due to gyration, and $\ogyr = 2\pi \fgyr$ and $\oext = 2\pi \fext$. The spectrum analyzer measures the power given by
\begin{equation}
\begin{split}
P(t) &= p_0 + p_1 \cos (\oext t) + p_2 \cos (2\oext t) + p_3 \cos (\ogyr t) \\ &+ p_4 \cos [(\ogyr + \oext )t] + p_4 \cos [(\ogyr - \oext )t ]) \\ &+ p_5 \cos [(\ogyr + 2\oext )t ] + p_5 \cos [(\ogyr - 2\oext )t ],
\end{split}
\end{equation}
where $p_0 = R_0 ( \Idc^2 + \iac^2 / 2 )$, $p_1 = 2 R_0 \Idc \iac$, $p_2 = R_0 \iac^2 /2 $, $p_3 = \Delta R ( \Idc^2 + \iac^2 / 2 )$, $p_4 = \Delta R \Idc \iac$, and $p_5 = \Delta R \iac^2 /4$. A number of these frequencies are visible experimentally, namely $\fext$, $2\fext$, $\fgyr$, and $\fgyr + \fext$. Some of the other frequencies are not clearly visible experimentally, due to their low intensity. However, they are much more visible in the simulation data, which can be seen in Fig.~\ref{fig:allforc}(d).

For the micromagnetics simulations presented in Fig.~\ref{fig:allforc}(d), we considered a DC current of $\Idc = 10$ mA which leads to a gyration frequency of 200 MHz. To study higher AC currents than those attainable experimentally, and to better visualize the modulation sidebands, we considered an AC current $\iac = 1$~mA that was swept from 100 to 900 MHz. For $\fext < \fgyr$, we can clearly observe modulation effects in the power spectra, where the $f = \fext$ signal is accompanied by sidebands not only at $f = \fgyr + \fext$ and $f = \fgyr - \fext$, but also at $f = 2\fgyr - \fext$ and to a lesser extent at $f = 3\fgyr - \fext$. We can also slightly see $2\fext - \fgyr$ or $\fgyr - 2\fext$ sidebands. As $\fext$ enters the locking window, frequency entrainment can clearly be observed over a range of 25 MHz in which the gyration frequency is controlled by the frequency of $\iac$. This entrainment is also visible in the first harmonic, where a segment with $f = 2\fext$ is visible in the locking window. Phase locking is also observable at $\fext = 3\fgyr$ and $\fext = 4\fgyr$, where at each harmonic the entrainment of the gyration frequency varies like $\fgyr = \fext/(n+1)$ with $n$ denoting the $n^\textrm{th}$ harmonic. This is accompanied by clear modulation signals at $f = (n+1)\fgyr - \fext$, which are most visible in the frequency range below the gyration frequency $f < \fgyr$. While most of these frequencies correspond to those predicted by the simple model, we see additional contributions in the experimental spectra. These can be understood as higher order modulation effects. The simulations results are similar to the phase-locking phenomena observed in nanopillar vortex oscillators.

%\section{Phase-locking and modulation}
\subsection{Forcing in the commensurate regime}
We now examine the effects of current modulation in the commensurate regime, where periodic core reversal occurs at a rate that is an integer fraction of the gyration frequency. The experimental spectra are presented in Fig.~\ref{fig:allforc}(b). The operating conditions consist of a DC current of $\Idc = 15$ mA, which in one experiment leads to a gyration frequency of $\fgyr = 410$ MHz, as shown in Fig.~\ref{fig:allforc}(b). In Fig.~\ref{fig:allforc}(b), we can see that the phase-locking and modulation patterns are similar to the previous case in which the dynamics comprises pure gyration, though these phenomena are more visible and modulation occurs on a larger range in the commensurate regime. The important difference here is that the external signal now modulates two distinct processes, the gyration and the periodic core reversal, where the frequency for the latter is denoted by $\fcr$.

Besides phase-locking at $\fext=\fgyr$, we also find evidence of entrainment when the external signal crosses one of the modulation sidebands due to vortex core reversal, namely, at $\fext = \fgyr \pm \fcr$, as shown in Fig.~\ref{fig:allforc}(b). Phase-locking of modulation sidebands and fractional synchronization are phenomena that have already been reported in previous studies on STNOs~\cite{singh_integer_2017}, though it is in a feedback loop, and not related to core reversal.

The spectral features with constant frequencies in Fig.~\ref{fig:allforc}(b), i.e., which are independent of $\fext$, can be expressed as linear combinations of the gyration frequency $\fgyr$ and the core reversal frequency $\fcr$. These are the natural frequencies of the vortex dynamics. A similar spectrum with natural frequencies is given in Fig.~\ref{fig:spectrajmap}(b), without any external forcing. For instance, when $\fgyr = 4\fcr$, one of the natural frequencies can be the sum of the first sideband of the gyration frequency (i.e., $\fgyr + \fcr$) and of the third sideband of the second harmonic of the gyration frequency (i.e., $2\fgyr - 3\fcr$). Since the dynamics are in the commensurate regime, we can write $\fgyr = a \fcr$ with integer $a$, the commensurate ratio between these two frequencies [$a=4$ in Figs.~\ref{fig:allforc}(b) and \ref{fig:allforc}(e)], which allows us to express the $k^\textrm{th}$ natural frequency $\fri$ simply as
\begin{equation}
    \fri = k \fcr ,
\end{equation}
with $i$ being an integer constant. Therefore, we can write $f_1 = \fcr$, $f_2 = 2\fcr$, $f_3 = 3\fcr = \fgyr - \fcr$, $f_4 = 4\fcr = \fgyr$, etc. We observe then that the external signal modulates all the natural frequencies $\fri$ to a certain degree. This simpler description of the system enables us to reuse the simple model, previously discussed for the pure gyration regime, for the commensurate regime,
\begin{equation}
\begin{split}
r(t) &= R_0 + \sumi \dri \cos (\oi t), \\
i(t) &= \Idc + \iac \cos (\oext t),
\end{split}
\end{equation}
where $R_0$, $\Delta R$, $\oext$, and $\oi$ are defined as in the previous section. Here, $\omega_1 = 2\pi \fcr$, $\omega_2 = 2\pi 2\fcr$, $\omega_3 = 2\pi 3\fcr$, $\omega_4 = \fgyr$ [due to the 1/4 ratio between $\fcr$ and $\fgyr$ in Fig.~\ref{fig:allforc}(b)]. The measured power is therefore
\begin{equation}
\begin{split}
P(t) &= p_0 + p_1 \cos (\oext t) + p_2 \cos (2\oext t) + \sumi p_\mathrm{3,k} \cos (\oi t) \\ &+ \sumi p_\mathrm{4,k} \cos [(\oi + \oext )t] + \sumi p_\mathrm{4,k} \cos [(\oi - \oext )t] \\ &+ \sumi p_\mathrm{5,k} \cos [(\oi + 2\oext )t] + \sumi p_\mathrm{5,k} \cos [(\oi - 2\oext )t],
\end{split}
\end{equation}
where $p_0 = R_0 ( \Idc^2 + \iac^2 / 2)$, $p_1 = 2 R_0 \Idc \iac$, $p_2 = R_0 \iac^2 /2$, $p_\mathrm{3,k} = \dri ( \Idc^2 + \iac^2 / 2)$, $p_\mathrm{4,k} = \dri \Idc \iac$, and $p_\mathrm{5,k} = \dri \iac^2 /4$. We can now compare this simple model to the frequencies exhibited in Fig.~\ref{fig:allforc}(b). Ascending branches correspond to the $f = \fri + \fext$ frequencies, while descending branches correspond to the $f = \fri - \fext$ sidebands. Only the first order of modulation is visible, such that no sidebands $f = \fri \pm n \fext$, with $n$ denoting the $n^\textrm{th}$ order, appear. No signals collinear to the harmonics of the external signal appear.

Phase-locking, like in the pure gyration regime, occurs when $\fext$ is equal to a multiple of $\fgyr$. But in this commensurate regime, it can occur also for any natural frequency of the vortex. When such a phase-locking occurs, it is also visible on the other natural frequencies of the vortex, at a higher or smaller frequency. Here, fractional synchronization is possible. However, $\fext$ can cross some natural frequency $\fri$ without inducing phase-locking. The reason why some natural frequencies are more likely to be phase-locked remains unknown, though we can at least say that frequencies like $\fgyr$, $\fcr$ and their harmonics are more likely to respond to external excitation than any other $\fri$. It should be noticed that core reversal corresponds to a square signal that contains only odd harmonics. Therefore, we see phase-locking at $\fcr$, $3\fcr$, $5\fcr$ but not at $2\fcr$, for instance.

Simulation in Fig.~\ref{fig:allforc}(e) exhibits a similar behavior than experimental curves, showing nonetheless a regime change which we did not observe in our measurements, but which have been shown on a different device \cite{du_hamel_de_milly_manipulation_2017}. This effect is slightly visible in Fig.~\ref{fig:allforc}(d), but is wider in Fig.~\ref{fig:allforc}(e): the system changes from a commensurate to an incommensurate regime, mainly around the phase-locking region. Indeed, between 260 and 310 MHz, and between 350 and 460 MHz, there are no phase-locking or commensurate regimes: we don't see any sharp peaks but rather a diffuse signal, which is characteristic of the incommensurate regime. Such a regime modification will be described with more details in Sec.~\ref{part5} and Fig.~\ref{fig:trajdt}.

Over a frequency range of approximatively 700~MHz, the $\fext$ signal appears stronger in Fig.~\ref{fig:allforc}(e) [indeed, it is also the case for Figs.~\ref{fig:allforc}(d) and \ref{fig:allforc}(f)]. This seems to indicate that for $\fext = \fgyr$, the influence of $\fext$ over the vortex decreases, and therefore less energy is transferred to it, leading to a more intense $\fext$.

\subsection{Incommensurate states}

When the vortex exhibits a chaotic behavior, the appearance of the PSD map changes in terms of phase-locking and modulation. Indeed, in Fig.~\ref{fig:allforc}(c), we apply a DC current of $\Idc = 16.7$ mA, leading to a gyration frequency of $\fgyr = 520$ MHz. We still apply an AC current of $\iac = 0.3$ mA. We can see that there is no phase-locking and modulation when the external and a vortex frequencies are incommensurate. This indicates that a chaotic behavior prevents phase-locking and modulation of such oscillators. Such a result is a consequence of the Kolmogorov-Arnold-Moser theory on dynamical systems~\cite{kolmogorov_conservation_1954, arnold_proof_1963, moser_invariant_1962}: an incommensurate regime is more robust to small perturbations. Indeed, a chaotic oscillator emits wider band frequencies, and therefore cannot be synchronized to a single frequency. Incommensurate states, which appear in chaotic regimes in vortex nanocontact oscillators~\cite{petit-watelot_commensurability_2012}, are less subject to phase-locking. Core reversal is aperiodic in a chaotic incommensurate regime. Therefore, a periodic external forcing barely induces locking of vortex frequencies. However, it might be possible that increasing the coupling strength between the chaotic oscillator and the external signal, in other terms, a higher AC current, induces phase-locking and modulation, making chaotic regime oscillators have a similar behavior to steady oscillation or core-reversal regime oscillators.

Indeed, in Fig.~\ref{fig:allforc}(f), where a higher AC current sent into the device is simulated, we can see a 30-MHz phase-locking range between $\fext$ and $\fgyr$. Modulation sidebands are also visible at low frequency.

%%
%	Section
%%
\section{Impact of current modulation on core reversal\label{part5}}

\begin{figure}
\centering\includegraphics[width=8cm]{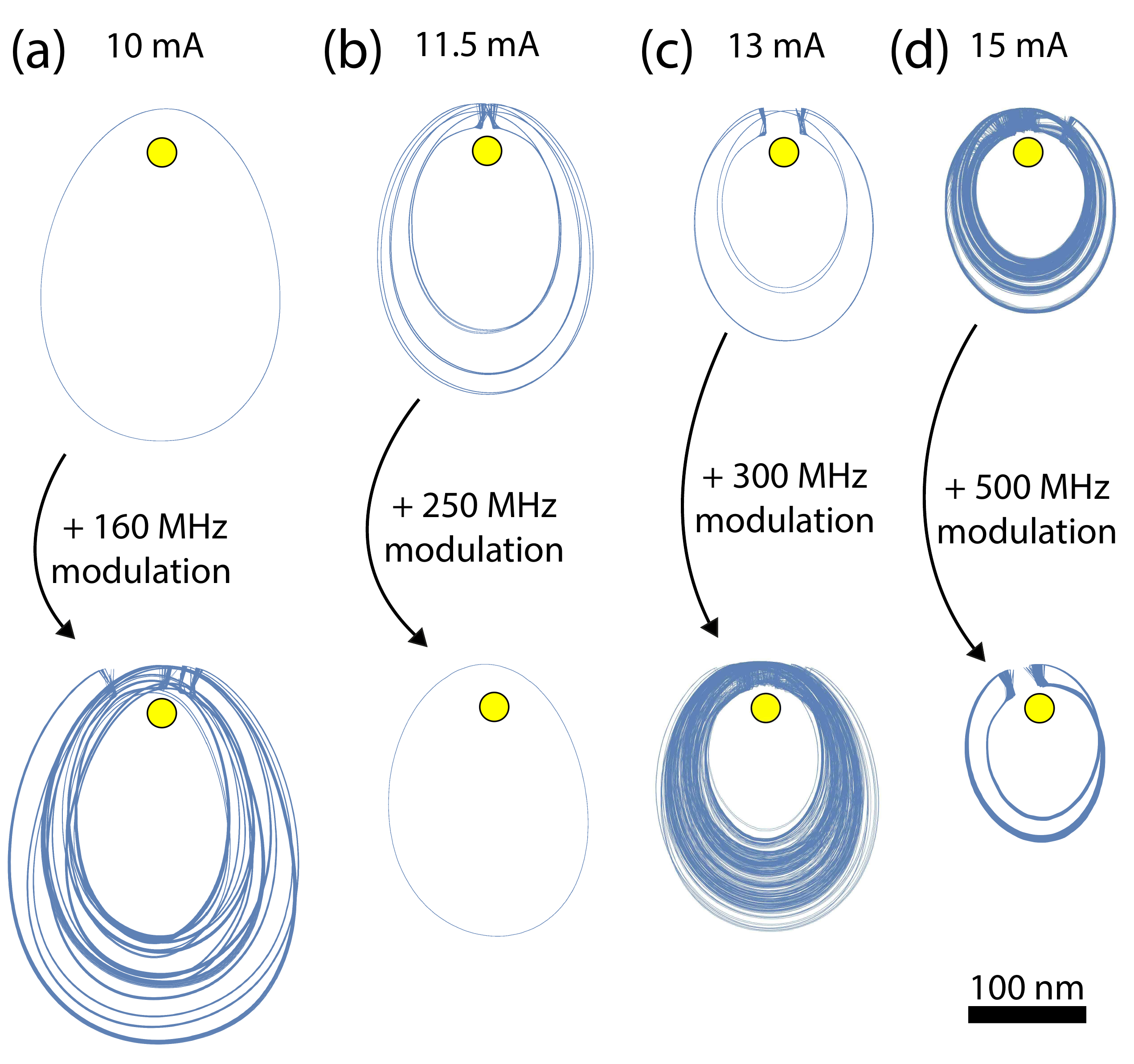}
\caption{Role of current modulation on vortex core reversal. Simulated trajectories, top ones correspond to the motion under DC currents alone, while bottom ones correspond to the addition of current modulation (at the frequencies indicated). (a,b,c,d) correspond to different operating conditions, with a given DC current.}
\label{fig:trajdt}
\end{figure}

To better understand the simulated spectra presented in Figs.~\ref{fig:allforc}(d), \ref{fig:allforc}(e), and \ref{fig:allforc}(f), we examine how the trajectories of the vortex core change when the current modulation is present. The trajectories can be classified into four  categories: a fixed point (no gyration), a limit cycle representing steady state gyration, a limit cycle with core reversal, and a chaotic attractor~\cite{devolder_chaos_2019}. In Fig.~\ref{fig:trajdt}, we illustrate examples of trajectories in the steady state gyration regime [Fig.~\ref{fig:trajdt}(a)], the commensurate state [Figs.~\ref{fig:trajdt}(b) and \ref{fig:trajdt}(c)], and the incommensurate or chaotic regime [Fig.~\ref{fig:trajdt}(d)].

Current modulation can change the oscillation regime. Figure~\ref{fig:trajdt}(a) shows transitions between the steady-state gyration toward a core-reversal state ($\Idc = 10$ mA, $\fext = 160$ MHz), and the opposite transition from a commensurate state toward steady-state gyration ($\Idc = 11.5$ mA, $\fext = 250$ MHz) can be seen in Fig.~\ref{fig:trajdt}(b). We also observe modulation-induced transitions back and forth between the commensurate and incommensurate states, which are shown in Fig.~\ref{fig:trajdt}(c) for $\Idc = 13$ mA, $\fext = 300$ MHz and in Fig.~\ref{fig:trajdt}(d) for $\Idc = 15$ mA, $\fext = 500$ MHz. This indicates that the conditions for core reversal can be suppressed or delayed as a result of the current modulation.

\begin{figure}
\centering\includegraphics[width=8.5cm]{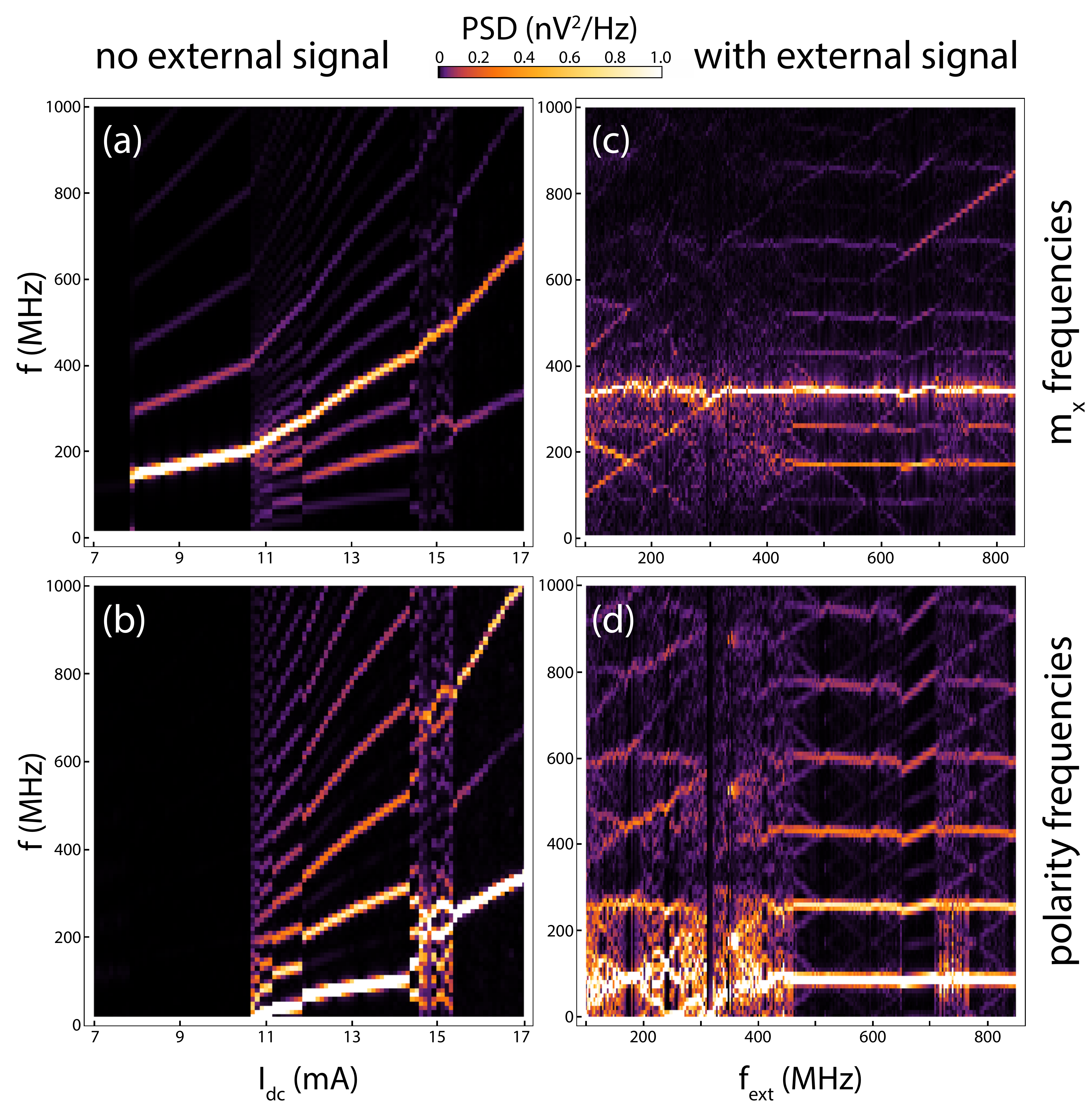}
\caption{ PSD maps of magnetoresistance signal [(a), (c)] and polarity frequencies [(b), (d)] while a DC current sweep without external signal [(a), (b)] or while an external frequency sweep with 13 mA DC current [(c), (d)].}
\label{fig:coremz}
\end{figure}

Let us now discuss these modulation-induced transitions in more detail. Figure~\ref{fig:coremz} illustrates the simulated power spectra of the magnetoresistance signal and the vortex core polarity, $p$, as a function of the DC current $\Idc$ and the modulation frequency $\fext$. 
In the absence of forcing, we can observe two clear thresholds for oscillations as the current is increased, one for the onset of steady-state gyration at around 8 mA [Fig.~\ref{fig:coremz}(a)] and the other for the onset of periodic core reversal at around 10.5 mA [Fig.~\ref{fig:coremz}(b)]. These figures clearly demonstrate that core reversal is at the origin of the intrinsic modulation that appears above 10.5 mA. Moreover, we can see that in the modulation regime, the spectra for the magnetoresistance and core polarization oscillations differ: only odd harmonics of $\fcr$ are visible in the PSD of the core polarization, whereas all harmonics of $\fcr$ are visible in the PSD of the magnetoresistance variations. Because the core polarity signal closely resembles a square wave, its Fourier series only contains odd harmonics. If there is jitter in the reversal events, even harmonics also might appear. On the other hand, the power spectrum of the magnetoresistance comprises the vortex gyration, which provides the dominant term in the PSD, along with the core reversal and the different orders of modulation between these two frequencies. This indicates that spectral lines such as $f = 2 \fcr$ at $\Idc \approx 15$ mA originates from second-order modulation processes between the gyration and core-reversal frequencies because the core reversal signal on its own cannot produce an even $\fcr$ harmonic.

We now present a similar analysis for a fixed current of 13 mA, where the modulation frequency is swept from 100 to 850 MHz [Figs.~\ref{fig:coremz}(c) and \ref{fig:coremz}(d)]. Similar modulation effects as those described previously are observed, except around the phase-locking region where the core-reversal frequency $\fcr$ decreases [Fig.~\ref{fig:coremz}(d)]. This is a consequence of the core reversal being impeded by the modulation, which leads to longer intervals between reversal events. As such, we can observe that the modulation frequencies disappear around the 1:1 locking in the power spectrum of the magnetoresistance variations [Fig.~\ref{fig:coremz}(c)].

\section{Discussion and concluding remarks}

We have performed a detailed experimental and numerical study of the role of current modulation on the vortex dynamics in magnetic nanocontact oscillators. These oscillators can possess two intrinsic modes, which can coexist: steady-state gyration around the nanocontact and periodic core reversal, which are characterized by the frequencies $\fgyr$ and $\fcr$, respectively. We have shown how modulation in the applied current, which affects both the Zeeman potential and the spin-transfer torques, influence these regimes. In particular, the modulation of both $\fgyr$ and $\fcr$ by the external frequency $\fext$ can lead to rich intermodulation spectra. External modulation can also lead to transitions between the natural oscillation regimes, namely simple gyration, commensurate, and chaotic states.

Beyond phase-locking applications such as spectrum analysis~\cite{louis_ultra-fast_2018}, we suggest that the nanocontact vortex oscillator might also be suitable for neuro-inspired applications~\cite{romera_vowel_2018, tsunegi_physical_2019} since it offers a rich variety of oscillatory modes that can be harnessed with external modulation. In terms of chaos-based information processing, this study sheds light on how unstable periodic orbits might be target using external modulation. It is indeed a first step in chaos control in nanocontact vortex oscillators. Another step would be to prove experimentally what has been seen in simulation, namely, the regime charge and phase-locking of chaos at higher forcing strength.

\begin{acknowledgments}
The authors thank S. Girod and C. Deranlot for assistance in sample fabrication, and J.-P. Adam and D. Rontani for fruitful discussions. This work is supported by the Agence Nationale de la Recherche (France) as part of the Investissements d'Avenir program (LabEx NanoSaclay, Reference No. ANR-10-LABX-0035) and under Contract No. ANR-17-CE24-0008 (CHIPMuNCS). J.L. acknowledges financial support from the EOBE doctoral school of Universit{\'e} Paris-Saclay. M.-W.Y. acknowledges financial support from the Horizon2020 Research Framework Programme of the European Commission under Contract No. 751344 (CHAOSPIN).
\end{acknowledgments}

\bibliography{main} %Pas d'espaces dans le nom de la biblio

\begin{thebibliography}{63}%
\makeatletter
\providecommand \@ifxundefined [1]{%
 \@ifx{#1\undefined}
}%
\providecommand \@ifnum [1]{%
 \ifnum #1\expandafter \@firstoftwo
 \else \expandafter \@secondoftwo
 \fi
}%
\providecommand \@ifx [1]{%
 \ifx #1\expandafter \@firstoftwo
 \else \expandafter \@secondoftwo
 \fi
}%
\providecommand \natexlab [1]{#1}%
\providecommand \enquote  [1]{``#1''}%
\providecommand \bibnamefont  [1]{#1}%
\providecommand \bibfnamefont [1]{#1}%
\providecommand \citenamefont [1]{#1}%
\providecommand \href@noop [0]{\@secondoftwo}%
\providecommand \href [0]{\begingroup \@sanitize@url \@href}%
\providecommand \@href[1]{\@@startlink{#1}\@@href}%
\providecommand \@@href[1]{\endgroup#1\@@endlink}%
\providecommand \@sanitize@url [0]{\catcode `\\12\catcode `\$12\catcode
  `\&12\catcode `\#12\catcode `\^12\catcode `\_12\catcode `\%12\relax}%
\providecommand \@@startlink[1]{}%
\providecommand \@@endlink[0]{}%
\providecommand \url  [0]{\begingroup\@sanitize@url \@url }%
\providecommand \@url [1]{\endgroup\@href {#1}{\urlprefix }}%
\providecommand \urlprefix  [0]{URL }%
\providecommand \Eprint [0]{\href }%
\providecommand \doibase [0]{http://dx.doi.org/}%
\providecommand \selectlanguage [0]{\@gobble}%
\providecommand \bibinfo  [0]{\@secondoftwo}%
\providecommand \bibfield  [0]{\@secondoftwo}%
\providecommand \translation [1]{[#1]}%
\providecommand \BibitemOpen [0]{}%
\providecommand \bibitemStop [0]{}%
\providecommand \bibitemNoStop [0]{.\EOS\space}%
\providecommand \EOS [0]{\spacefactor3000\relax}%
\providecommand \BibitemShut  [1]{\csname bibitem#1\endcsname}%
\let\auto@bib@innerbib\@empty
%</preamble>
\bibitem [{\citenamefont
  {Slonczewski}(1996)}]{slonczewski_current-driven_1996}%
  \BibitemOpen
  \bibfield  {author} {\bibinfo {author} {\bibfnamefont {J.}~\bibnamefont
  {Slonczewski}},\ }\href {\doibase 10.1016/0304-8853(96)00062-5} {\bibfield
  {journal} {\bibinfo  {journal} {Journal of Magnetism and Magnetic Materials}\
  }\textbf {\bibinfo {volume} {159}},\ \bibinfo {pages} {L1} (\bibinfo {year}
  {1996})}\BibitemShut {NoStop}%
\bibitem [{\citenamefont {Berger}(1996)}]{berger_emission_1996}%
  \BibitemOpen
  \bibfield  {author} {\bibinfo {author} {\bibfnamefont {L.}~\bibnamefont
  {Berger}},\ }\href {\doibase 10.1103/PhysRevB.54.9353} {\bibfield  {journal}
  {\bibinfo  {journal} {Physical Review B}\ }\textbf {\bibinfo {volume} {54}},\
  \bibinfo {pages} {9353} (\bibinfo {year} {1996})}\BibitemShut {NoStop}%
\bibitem [{\citenamefont {Kiselev}\ \emph {et~al.}(2003)\citenamefont
  {Kiselev}, \citenamefont {Sankey}, \citenamefont {Krivorotov}, \citenamefont
  {Emley}, \citenamefont {Schoelkopf}, \citenamefont {Buhrman},\ and\
  \citenamefont {Ralph}}]{kiselev_microwave_2003}%
  \BibitemOpen
  \bibfield  {author} {\bibinfo {author} {\bibfnamefont {S.~I.}\ \bibnamefont
  {Kiselev}}, \bibinfo {author} {\bibfnamefont {J.~C.}\ \bibnamefont {Sankey}},
  \bibinfo {author} {\bibfnamefont {I.~N.}\ \bibnamefont {Krivorotov}},
  \bibinfo {author} {\bibfnamefont {N.~C.}\ \bibnamefont {Emley}}, \bibinfo
  {author} {\bibfnamefont {R.~J.}\ \bibnamefont {Schoelkopf}}, \bibinfo
  {author} {\bibfnamefont {R.~A.}\ \bibnamefont {Buhrman}}, \ and\ \bibinfo
  {author} {\bibfnamefont {D.~C.}\ \bibnamefont {Ralph}},\ }\href {\doibase
  10.1038/nature01967} {\bibfield  {journal} {\bibinfo  {journal} {Nature}\
  }\textbf {\bibinfo {volume} {425}},\ \bibinfo {pages} {380} (\bibinfo {year}
  {2003})}\BibitemShut {NoStop}%
\bibitem [{\citenamefont {Slavin}\ and\ \citenamefont
  {Tiberkevich}(2009)}]{slavin_nonlinear_2009}%
  \BibitemOpen
  \bibfield  {author} {\bibinfo {author} {\bibfnamefont {A.}~\bibnamefont
  {Slavin}}\ and\ \bibinfo {author} {\bibfnamefont {V.}~\bibnamefont
  {Tiberkevich}},\ }\href {\doibase 10.1109/TMAG.2008.2009935} {\bibfield
  {journal} {\bibinfo  {journal} {IEEE Transactions on Magnetics}\ }\textbf
  {\bibinfo {volume} {45}},\ \bibinfo {pages} {1875} (\bibinfo {year}
  {2009})}\BibitemShut {NoStop}%
\bibitem [{\citenamefont {Bonetti}\ \emph {et~al.}(2015)\citenamefont
  {Bonetti}, \citenamefont {Kukreja}, \citenamefont {Chen}, \citenamefont
  {Maci{\`a}}, \citenamefont {Hern{\`a}ndez}, \citenamefont {Eklund},
  \citenamefont {Backes}, \citenamefont {Frisch}, \citenamefont {Katine},
  \citenamefont {Malm}, \citenamefont {Urazhdin}, \citenamefont {Kent},
  \citenamefont {St{\"o}hr}, \citenamefont {Ohldag},\ and\ \citenamefont
  {D{\"u}rr}}]{bonetti_direct_2015}%
  \BibitemOpen
  \bibfield  {author} {\bibinfo {author} {\bibfnamefont {S.}~\bibnamefont
  {Bonetti}}, \bibinfo {author} {\bibfnamefont {R.}~\bibnamefont {Kukreja}},
  \bibinfo {author} {\bibfnamefont {Z.}~\bibnamefont {Chen}}, \bibinfo {author}
  {\bibfnamefont {F.}~\bibnamefont {Maci{\`a}}}, \bibinfo {author}
  {\bibfnamefont {J.~M.}\ \bibnamefont {Hern{\`a}ndez}}, \bibinfo {author}
  {\bibfnamefont {A.}~\bibnamefont {Eklund}}, \bibinfo {author} {\bibfnamefont
  {D.}~\bibnamefont {Backes}}, \bibinfo {author} {\bibfnamefont
  {J.}~\bibnamefont {Frisch}}, \bibinfo {author} {\bibfnamefont
  {J.}~\bibnamefont {Katine}}, \bibinfo {author} {\bibfnamefont
  {G.}~\bibnamefont {Malm}}, \bibinfo {author} {\bibfnamefont {S.}~\bibnamefont
  {Urazhdin}}, \bibinfo {author} {\bibfnamefont {A.~D.}\ \bibnamefont {Kent}},
  \bibinfo {author} {\bibfnamefont {J.}~\bibnamefont {St{\"o}hr}}, \bibinfo
  {author} {\bibfnamefont {H.}~\bibnamefont {Ohldag}}, \ and\ \bibinfo {author}
  {\bibfnamefont {H.~A.}\ \bibnamefont {D{\"u}rr}},\ }\href {\doibase
  10.1038/ncomms9889} {\bibfield  {journal} {\bibinfo  {journal} {Nature
  Communications}\ }\textbf {\bibinfo {volume} {6}},\ \bibinfo {pages} {8889}
  (\bibinfo {year} {2015})}\BibitemShut {NoStop}%
\bibitem [{\citenamefont {Kreissig}\ \emph {et~al.}(2017)\citenamefont
  {Kreissig}, \citenamefont {Lebrun}, \citenamefont {Protze}, \citenamefont
  {Merazzo}, \citenamefont {Hem}, \citenamefont {Vila}, \citenamefont
  {Ferreira}, \citenamefont {Cyrille}, \citenamefont {Ellinger}, \citenamefont
  {Cros}, \citenamefont {Ebels},\ and\ \citenamefont
  {Bortolotti}}]{kreissig_vortex_2017}%
  \BibitemOpen
  \bibfield  {author} {\bibinfo {author} {\bibfnamefont {M.}~\bibnamefont
  {Kreissig}}, \bibinfo {author} {\bibfnamefont {R.}~\bibnamefont {Lebrun}},
  \bibinfo {author} {\bibfnamefont {F.}~\bibnamefont {Protze}}, \bibinfo
  {author} {\bibfnamefont {K.~J.}\ \bibnamefont {Merazzo}}, \bibinfo {author}
  {\bibfnamefont {J.}~\bibnamefont {Hem}}, \bibinfo {author} {\bibfnamefont
  {L.}~\bibnamefont {Vila}}, \bibinfo {author} {\bibfnamefont {R.}~\bibnamefont
  {Ferreira}}, \bibinfo {author} {\bibfnamefont {M.~C.}\ \bibnamefont
  {Cyrille}}, \bibinfo {author} {\bibfnamefont {F.}~\bibnamefont {Ellinger}},
  \bibinfo {author} {\bibfnamefont {V.}~\bibnamefont {Cros}}, \bibinfo {author}
  {\bibfnamefont {U.}~\bibnamefont {Ebels}}, \ and\ \bibinfo {author}
  {\bibfnamefont {P.}~\bibnamefont {Bortolotti}},\ }\href {\doibase
  10.1063/1.4976337} {\bibfield  {journal} {\bibinfo  {journal} {AIP Advances}\
  }\textbf {\bibinfo {volume} {7}},\ \bibinfo {pages} {056653} (\bibinfo {year}
  {2017})}\BibitemShut {NoStop}%
\bibitem [{\citenamefont {Tulapurkar}\ \emph {et~al.}(2005)\citenamefont
  {Tulapurkar}, \citenamefont {Suzuki}, \citenamefont {Fukushima},
  \citenamefont {Kubota}, \citenamefont {Maehara}, \citenamefont {Tsunekawa},
  \citenamefont {Djayaprawira}, \citenamefont {Watanabe},\ and\ \citenamefont
  {Yuasa}}]{tulapurkar_spin-torque_2005}%
  \BibitemOpen
  \bibfield  {author} {\bibinfo {author} {\bibfnamefont {A.~A.}\ \bibnamefont
  {Tulapurkar}}, \bibinfo {author} {\bibfnamefont {Y.}~\bibnamefont {Suzuki}},
  \bibinfo {author} {\bibfnamefont {A.}~\bibnamefont {Fukushima}}, \bibinfo
  {author} {\bibfnamefont {H.}~\bibnamefont {Kubota}}, \bibinfo {author}
  {\bibfnamefont {H.}~\bibnamefont {Maehara}}, \bibinfo {author} {\bibfnamefont
  {K.}~\bibnamefont {Tsunekawa}}, \bibinfo {author} {\bibfnamefont {D.~D.}\
  \bibnamefont {Djayaprawira}}, \bibinfo {author} {\bibfnamefont
  {N.}~\bibnamefont {Watanabe}}, \ and\ \bibinfo {author} {\bibfnamefont
  {S.}~\bibnamefont {Yuasa}},\ }\href {\doibase 10.1038/nature04207} {\bibfield
   {journal} {\bibinfo  {journal} {Nature}\ }\textbf {\bibinfo {volume}
  {438}},\ \bibinfo {pages} {339} (\bibinfo {year} {2005})}\BibitemShut
  {NoStop}%
\bibitem [{\citenamefont {Torrejon}\ \emph {et~al.}(2017)\citenamefont
  {Torrejon}, \citenamefont {Riou}, \citenamefont {Araujo}, \citenamefont
  {Tsunegi}, \citenamefont {Khalsa}, \citenamefont {Querlioz}, \citenamefont
  {Bortolotti}, \citenamefont {Cros}, \citenamefont {Yakushiji}, \citenamefont
  {Fukushima}, \citenamefont {Kubota}, \citenamefont {Yuasa}, \citenamefont
  {Stiles},\ and\ \citenamefont {Grollier}}]{torrejon_neuromorphic_2017}%
  \BibitemOpen
  \bibfield  {author} {\bibinfo {author} {\bibfnamefont {J.}~\bibnamefont
  {Torrejon}}, \bibinfo {author} {\bibfnamefont {M.}~\bibnamefont {Riou}},
  \bibinfo {author} {\bibfnamefont {F.~A.}\ \bibnamefont {Araujo}}, \bibinfo
  {author} {\bibfnamefont {S.}~\bibnamefont {Tsunegi}}, \bibinfo {author}
  {\bibfnamefont {G.}~\bibnamefont {Khalsa}}, \bibinfo {author} {\bibfnamefont
  {D.}~\bibnamefont {Querlioz}}, \bibinfo {author} {\bibfnamefont
  {P.}~\bibnamefont {Bortolotti}}, \bibinfo {author} {\bibfnamefont
  {V.}~\bibnamefont {Cros}}, \bibinfo {author} {\bibfnamefont {K.}~\bibnamefont
  {Yakushiji}}, \bibinfo {author} {\bibfnamefont {A.}~\bibnamefont
  {Fukushima}}, \bibinfo {author} {\bibfnamefont {H.}~\bibnamefont {Kubota}},
  \bibinfo {author} {\bibfnamefont {S.}~\bibnamefont {Yuasa}}, \bibinfo
  {author} {\bibfnamefont {M.~D.}\ \bibnamefont {Stiles}}, \ and\ \bibinfo
  {author} {\bibfnamefont {J.}~\bibnamefont {Grollier}},\ }\href {\doibase
  10.1038/nature23011} {\bibfield  {journal} {\bibinfo  {journal} {Nature}\
  }\textbf {\bibinfo {volume} {547}},\ \bibinfo {pages} {428} (\bibinfo {year}
  {2017})}\BibitemShut {NoStop}%
\bibitem [{\citenamefont {Romera}\ \emph {et~al.}(2018)\citenamefont {Romera},
  \citenamefont {Talatchian}, \citenamefont {Tsunegi}, \citenamefont {Araujo},
  \citenamefont {Cros}, \citenamefont {Bortolotti}, \citenamefont {Trastoy},
  \citenamefont {Yakushiji}, \citenamefont {Fukushima}, \citenamefont {Kubota},
  \citenamefont {Yuasa}, \citenamefont {Ernoult}, \citenamefont
  {Vodenicarevic}, \citenamefont {Hirtzlin}, \citenamefont {Locatelli},
  \citenamefont {Querlioz},\ and\ \citenamefont
  {Grollier}}]{romera_vowel_2018}%
  \BibitemOpen
  \bibfield  {author} {\bibinfo {author} {\bibfnamefont {M.}~\bibnamefont
  {Romera}}, \bibinfo {author} {\bibfnamefont {P.}~\bibnamefont {Talatchian}},
  \bibinfo {author} {\bibfnamefont {S.}~\bibnamefont {Tsunegi}}, \bibinfo
  {author} {\bibfnamefont {F.~A.}\ \bibnamefont {Araujo}}, \bibinfo {author}
  {\bibfnamefont {V.}~\bibnamefont {Cros}}, \bibinfo {author} {\bibfnamefont
  {P.}~\bibnamefont {Bortolotti}}, \bibinfo {author} {\bibfnamefont
  {J.}~\bibnamefont {Trastoy}}, \bibinfo {author} {\bibfnamefont
  {K.}~\bibnamefont {Yakushiji}}, \bibinfo {author} {\bibfnamefont
  {A.}~\bibnamefont {Fukushima}}, \bibinfo {author} {\bibfnamefont
  {H.}~\bibnamefont {Kubota}}, \bibinfo {author} {\bibfnamefont
  {S.}~\bibnamefont {Yuasa}}, \bibinfo {author} {\bibfnamefont
  {M.}~\bibnamefont {Ernoult}}, \bibinfo {author} {\bibfnamefont
  {D.}~\bibnamefont {Vodenicarevic}}, \bibinfo {author} {\bibfnamefont
  {T.}~\bibnamefont {Hirtzlin}}, \bibinfo {author} {\bibfnamefont
  {N.}~\bibnamefont {Locatelli}}, \bibinfo {author} {\bibfnamefont
  {D.}~\bibnamefont {Querlioz}}, \ and\ \bibinfo {author} {\bibfnamefont
  {J.}~\bibnamefont {Grollier}},\ }\href {\doibase 10.1038/s41586-018-0632-y}
  {\bibfield  {journal} {\bibinfo  {journal} {Nature}\ }\textbf {\bibinfo
  {volume} {563}},\ \bibinfo {pages} {230} (\bibinfo {year}
  {2018})}\BibitemShut {NoStop}%
\bibitem [{\citenamefont {Tsunegi}\ \emph {et~al.}(2019)\citenamefont
  {Tsunegi}, \citenamefont {Taniguchi}, \citenamefont {Nakajima}, \citenamefont
  {Miwa}, \citenamefont {Yakushiji}, \citenamefont {Fukushima}, \citenamefont
  {Yuasa},\ and\ \citenamefont {Kubota}}]{tsunegi_physical_2019}%
  \BibitemOpen
  \bibfield  {author} {\bibinfo {author} {\bibfnamefont {S.}~\bibnamefont
  {Tsunegi}}, \bibinfo {author} {\bibfnamefont {T.}~\bibnamefont {Taniguchi}},
  \bibinfo {author} {\bibfnamefont {K.}~\bibnamefont {Nakajima}}, \bibinfo
  {author} {\bibfnamefont {S.}~\bibnamefont {Miwa}}, \bibinfo {author}
  {\bibfnamefont {K.}~\bibnamefont {Yakushiji}}, \bibinfo {author}
  {\bibfnamefont {A.}~\bibnamefont {Fukushima}}, \bibinfo {author}
  {\bibfnamefont {S.}~\bibnamefont {Yuasa}}, \ and\ \bibinfo {author}
  {\bibfnamefont {H.}~\bibnamefont {Kubota}},\ }\href {\doibase
  10.1063/1.5081797} {\bibfield  {journal} {\bibinfo  {journal} {Applied
  Physics Letters}\ }\textbf {\bibinfo {volume} {114}},\ \bibinfo {pages}
  {164101} (\bibinfo {year} {2019})}\BibitemShut {NoStop}%
\bibitem [{\citenamefont {Rippard}\ \emph {et~al.}(2005)\citenamefont
  {Rippard}, \citenamefont {Pufall}, \citenamefont {Kaka}, \citenamefont
  {Silva}, \citenamefont {Russek},\ and\ \citenamefont
  {Katine}}]{rippard_injection_2005}%
  \BibitemOpen
  \bibfield  {author} {\bibinfo {author} {\bibfnamefont {W.~H.}\ \bibnamefont
  {Rippard}}, \bibinfo {author} {\bibfnamefont {M.~R.}\ \bibnamefont {Pufall}},
  \bibinfo {author} {\bibfnamefont {S.}~\bibnamefont {Kaka}}, \bibinfo {author}
  {\bibfnamefont {T.~J.}\ \bibnamefont {Silva}}, \bibinfo {author}
  {\bibfnamefont {S.~E.}\ \bibnamefont {Russek}}, \ and\ \bibinfo {author}
  {\bibfnamefont {J.~A.}\ \bibnamefont {Katine}},\ }\href {\doibase
  10.1103/PhysRevLett.95.067203} {\bibfield  {journal} {\bibinfo  {journal}
  {Physical Review Letters}\ }\textbf {\bibinfo {volume} {95}},\ \bibinfo
  {pages} {067203} (\bibinfo {year} {2005})}\BibitemShut {NoStop}%
\bibitem [{\citenamefont {Puliafito}\ \emph {et~al.}(2014)\citenamefont
  {Puliafito}, \citenamefont {Consolo}, \citenamefont {Lopez-Diaz},\ and\
  \citenamefont {Azzerboni}}]{puliafito_synchronization_2014}%
  \BibitemOpen
  \bibfield  {author} {\bibinfo {author} {\bibfnamefont {V.}~\bibnamefont
  {Puliafito}}, \bibinfo {author} {\bibfnamefont {G.}~\bibnamefont {Consolo}},
  \bibinfo {author} {\bibfnamefont {L.}~\bibnamefont {Lopez-Diaz}}, \ and\
  \bibinfo {author} {\bibfnamefont {B.}~\bibnamefont {Azzerboni}},\ }\href
  {\doibase 10.1016/j.physb.2013.09.042} {\bibfield  {journal} {\bibinfo
  {journal} {Physica B: Condensed Matter}\ }\bibinfo {series} {9th
  {International} {Symposium} on {Hysteresis} {Modeling} and {Micromagnetics}
  ({HMM} 2013)},\ \textbf {\bibinfo {volume} {435}},\ \bibinfo {pages} {44}
  (\bibinfo {year} {2014})}\BibitemShut {NoStop}%
\bibitem [{\citenamefont {Lebrun}\ \emph {et~al.}(2015)\citenamefont {Lebrun},
  \citenamefont {Jenkins}, \citenamefont {Dussaux}, \citenamefont {Locatelli},
  \citenamefont {Tsunegi}, \citenamefont {Grimaldi}, \citenamefont {Kubota},
  \citenamefont {Bortolotti}, \citenamefont {Yakushiji}, \citenamefont
  {Grollier}, \citenamefont {Fukushima}, \citenamefont {Yuasa},\ and\
  \citenamefont {Cros}}]{lebrun_understanding_2015}%
  \BibitemOpen
  \bibfield  {author} {\bibinfo {author} {\bibfnamefont {R.}~\bibnamefont
  {Lebrun}}, \bibinfo {author} {\bibfnamefont {A.}~\bibnamefont {Jenkins}},
  \bibinfo {author} {\bibfnamefont {A.}~\bibnamefont {Dussaux}}, \bibinfo
  {author} {\bibfnamefont {N.}~\bibnamefont {Locatelli}}, \bibinfo {author}
  {\bibfnamefont {S.}~\bibnamefont {Tsunegi}}, \bibinfo {author} {\bibfnamefont
  {E.}~\bibnamefont {Grimaldi}}, \bibinfo {author} {\bibfnamefont
  {H.}~\bibnamefont {Kubota}}, \bibinfo {author} {\bibfnamefont
  {P.}~\bibnamefont {Bortolotti}}, \bibinfo {author} {\bibfnamefont
  {K.}~\bibnamefont {Yakushiji}}, \bibinfo {author} {\bibfnamefont
  {J.}~\bibnamefont {Grollier}}, \bibinfo {author} {\bibfnamefont
  {A.}~\bibnamefont {Fukushima}}, \bibinfo {author} {\bibfnamefont
  {S.}~\bibnamefont {Yuasa}}, \ and\ \bibinfo {author} {\bibfnamefont
  {V.}~\bibnamefont {Cros}},\ }\href {\doibase 10.1103/PhysRevLett.115.017201}
  {\bibfield  {journal} {\bibinfo  {journal} {Physical Review Letters}\
  }\textbf {\bibinfo {volume} {115}},\ \bibinfo {pages} {017201} (\bibinfo
  {year} {2015})}\BibitemShut {NoStop}%
\bibitem [{\citenamefont {Gopal}\ \emph {et~al.}(2019)\citenamefont {Gopal},
  \citenamefont {Subash}, \citenamefont {Chandrasekar},\ and\ \citenamefont
  {Lakshmanan}}]{gopal_phase_2019}%
  \BibitemOpen
  \bibfield  {author} {\bibinfo {author} {\bibfnamefont {R.}~\bibnamefont
  {Gopal}}, \bibinfo {author} {\bibfnamefont {B.}~\bibnamefont {Subash}},
  \bibinfo {author} {\bibfnamefont {V.~K.}\ \bibnamefont {Chandrasekar}}, \
  and\ \bibinfo {author} {\bibfnamefont {M.}~\bibnamefont {Lakshmanan}},\
  }\href {\doibase 10.1109/TMAG.2019.2908954} {\bibfield  {journal} {\bibinfo
  {journal} {IEEE Transactions on Magnetics}\ }\textbf {\bibinfo {volume}
  {55}},\ \bibinfo {pages} {1} (\bibinfo {year} {2019})}\BibitemShut {NoStop}%
\bibitem [{\citenamefont {Pufall}\ \emph {et~al.}(2005)\citenamefont {Pufall},
  \citenamefont {Rippard}, \citenamefont {Kaka}, \citenamefont {Silva},\ and\
  \citenamefont {Russek}}]{pufall_frequency_2005}%
  \BibitemOpen
  \bibfield  {author} {\bibinfo {author} {\bibfnamefont {M.~R.}\ \bibnamefont
  {Pufall}}, \bibinfo {author} {\bibfnamefont {W.~H.}\ \bibnamefont {Rippard}},
  \bibinfo {author} {\bibfnamefont {S.}~\bibnamefont {Kaka}}, \bibinfo {author}
  {\bibfnamefont {T.~J.}\ \bibnamefont {Silva}}, \ and\ \bibinfo {author}
  {\bibfnamefont {S.~E.}\ \bibnamefont {Russek}},\ }\href {\doibase
  10.1063/1.1875762} {\bibfield  {journal} {\bibinfo  {journal} {Applied
  Physics Letters}\ }\textbf {\bibinfo {volume} {86}},\ \bibinfo {pages}
  {082506} (\bibinfo {year} {2005})}\BibitemShut {NoStop}%
\bibitem [{\citenamefont {Consolo}\ \emph {et~al.}(2010)\citenamefont
  {Consolo}, \citenamefont {Puliafito}, \citenamefont {Finocchio},
  \citenamefont {Lopez-Diaz}, \citenamefont {Zivieri}, \citenamefont
  {Giovannini}, \citenamefont {Nizzoli}, \citenamefont {Valenti},\ and\
  \citenamefont {Azzerboni}}]{consolo_combined_2010}%
  \BibitemOpen
  \bibfield  {author} {\bibinfo {author} {\bibfnamefont {G.}~\bibnamefont
  {Consolo}}, \bibinfo {author} {\bibfnamefont {V.}~\bibnamefont {Puliafito}},
  \bibinfo {author} {\bibfnamefont {G.}~\bibnamefont {Finocchio}}, \bibinfo
  {author} {\bibfnamefont {L.}~\bibnamefont {Lopez-Diaz}}, \bibinfo {author}
  {\bibfnamefont {R.}~\bibnamefont {Zivieri}}, \bibinfo {author} {\bibfnamefont
  {L.}~\bibnamefont {Giovannini}}, \bibinfo {author} {\bibfnamefont
  {F.}~\bibnamefont {Nizzoli}}, \bibinfo {author} {\bibfnamefont
  {G.}~\bibnamefont {Valenti}}, \ and\ \bibinfo {author} {\bibfnamefont
  {B.}~\bibnamefont {Azzerboni}},\ }\href {\doibase 10.1109/TMAG.2010.2046178}
  {\bibfield  {journal} {\bibinfo  {journal} {IEEE Transactions on Magnetics}\
  }\textbf {\bibinfo {volume} {46}},\ \bibinfo {pages} {3629} (\bibinfo {year}
  {2010})}\BibitemShut {NoStop}%
\bibitem [{\citenamefont {Muduli}\ \emph {et~al.}(2010)\citenamefont {Muduli},
  \citenamefont {Pogoryelov}, \citenamefont {Bonetti}, \citenamefont {Consolo},
  \citenamefont {Mancoff},\ and\ \citenamefont {{\r
  A}kerman}}]{muduli_nonlinear_2010}%
  \BibitemOpen
  \bibfield  {author} {\bibinfo {author} {\bibfnamefont {P.~K.}\ \bibnamefont
  {Muduli}}, \bibinfo {author} {\bibfnamefont {Y.}~\bibnamefont {Pogoryelov}},
  \bibinfo {author} {\bibfnamefont {S.}~\bibnamefont {Bonetti}}, \bibinfo
  {author} {\bibfnamefont {G.}~\bibnamefont {Consolo}}, \bibinfo {author}
  {\bibfnamefont {F.}~\bibnamefont {Mancoff}}, \ and\ \bibinfo {author}
  {\bibfnamefont {J.}~\bibnamefont {{\r A}kerman}},\ }\href {\doibase
  10.1103/PhysRevB.81.140408} {\bibfield  {journal} {\bibinfo  {journal}
  {Physical Review B}\ }\textbf {\bibinfo {volume} {81}},\ \bibinfo {pages}
  {140408} (\bibinfo {year} {2010})}\BibitemShut {NoStop}%
\bibitem [{\citenamefont {Pogoryelov}\ \emph {et~al.}(2011)\citenamefont
  {Pogoryelov}, \citenamefont {Muduli}, \citenamefont {Bonetti}, \citenamefont
  {Mancoff},\ and\ \citenamefont {{\r A}kerman}}]{pogoryelov_spin-torque_2011}%
  \BibitemOpen
  \bibfield  {author} {\bibinfo {author} {\bibfnamefont {Y.}~\bibnamefont
  {Pogoryelov}}, \bibinfo {author} {\bibfnamefont {P.~K.}\ \bibnamefont
  {Muduli}}, \bibinfo {author} {\bibfnamefont {S.}~\bibnamefont {Bonetti}},
  \bibinfo {author} {\bibfnamefont {F.}~\bibnamefont {Mancoff}}, \ and\
  \bibinfo {author} {\bibfnamefont {J.}~\bibnamefont {{\r A}kerman}},\ }\href
  {\doibase 10.1063/1.3588038} {\bibfield  {journal} {\bibinfo  {journal}
  {Applied Physics Letters}\ }\textbf {\bibinfo {volume} {98}},\ \bibinfo
  {pages} {192506} (\bibinfo {year} {2011})}\BibitemShut {NoStop}%
\bibitem [{\citenamefont {Martin}\ \emph {et~al.}(2013)\citenamefont {Martin},
  \citenamefont {Thirion}, \citenamefont {Hoarau}, \citenamefont {Baraduc},\
  and\ \citenamefont {Di{\'e}ny}}]{martin_tunability_2013}%
  \BibitemOpen
  \bibfield  {author} {\bibinfo {author} {\bibfnamefont {S.~Y.}\ \bibnamefont
  {Martin}}, \bibinfo {author} {\bibfnamefont {C.}~\bibnamefont {Thirion}},
  \bibinfo {author} {\bibfnamefont {C.}~\bibnamefont {Hoarau}}, \bibinfo
  {author} {\bibfnamefont {C.}~\bibnamefont {Baraduc}}, \ and\ \bibinfo
  {author} {\bibfnamefont {B.}~\bibnamefont {Di{\'e}ny}},\ }\href {\doibase
  10.1103/PhysRevB.88.024421} {\bibfield  {journal} {\bibinfo  {journal}
  {Physical Review B}\ }\textbf {\bibinfo {volume} {88}},\ \bibinfo {pages}
  {024421} (\bibinfo {year} {2013})}\BibitemShut {NoStop}%
\bibitem [{\citenamefont {Dussaux}\ \emph {et~al.}(2011)\citenamefont
  {Dussaux}, \citenamefont {Khvalkovskiy}, \citenamefont {Grollier},
  \citenamefont {Cros}, \citenamefont {Fukushima}, \citenamefont {Konoto},
  \citenamefont {Kubota}, \citenamefont {Yakushiji}, \citenamefont {Yuasa},
  \citenamefont {Ando},\ and\ \citenamefont {Fert}}]{dussaux_phase_2011}%
  \BibitemOpen
  \bibfield  {author} {\bibinfo {author} {\bibfnamefont {A.}~\bibnamefont
  {Dussaux}}, \bibinfo {author} {\bibfnamefont {A.~V.}\ \bibnamefont
  {Khvalkovskiy}}, \bibinfo {author} {\bibfnamefont {J.}~\bibnamefont
  {Grollier}}, \bibinfo {author} {\bibfnamefont {V.}~\bibnamefont {Cros}},
  \bibinfo {author} {\bibfnamefont {A.}~\bibnamefont {Fukushima}}, \bibinfo
  {author} {\bibfnamefont {M.}~\bibnamefont {Konoto}}, \bibinfo {author}
  {\bibfnamefont {H.}~\bibnamefont {Kubota}}, \bibinfo {author} {\bibfnamefont
  {K.}~\bibnamefont {Yakushiji}}, \bibinfo {author} {\bibfnamefont
  {S.}~\bibnamefont {Yuasa}}, \bibinfo {author} {\bibfnamefont
  {K.}~\bibnamefont {Ando}}, \ and\ \bibinfo {author} {\bibfnamefont
  {A.}~\bibnamefont {Fert}},\ }\href {\doibase 10.1063/1.3565159} {\bibfield
  {journal} {\bibinfo  {journal} {Applied Physics Letters}\ }\textbf {\bibinfo
  {volume} {98}},\ \bibinfo {pages} {132506} (\bibinfo {year}
  {2011})}\BibitemShut {NoStop}%
\bibitem [{\citenamefont {Hamadeh}\ \emph {et~al.}(2014)\citenamefont
  {Hamadeh}, \citenamefont {Locatelli}, \citenamefont {Naletov}, \citenamefont
  {Lebrun}, \citenamefont {de~Loubens}, \citenamefont {Grollier}, \citenamefont
  {Klein},\ and\ \citenamefont {Cros}}]{hamadeh_perfect_2014}%
  \BibitemOpen
  \bibfield  {author} {\bibinfo {author} {\bibfnamefont {A.}~\bibnamefont
  {Hamadeh}}, \bibinfo {author} {\bibfnamefont {N.}~\bibnamefont {Locatelli}},
  \bibinfo {author} {\bibfnamefont {V.~V.}\ \bibnamefont {Naletov}}, \bibinfo
  {author} {\bibfnamefont {R.}~\bibnamefont {Lebrun}}, \bibinfo {author}
  {\bibfnamefont {G.}~\bibnamefont {de~Loubens}}, \bibinfo {author}
  {\bibfnamefont {J.}~\bibnamefont {Grollier}}, \bibinfo {author}
  {\bibfnamefont {O.}~\bibnamefont {Klein}}, \ and\ \bibinfo {author}
  {\bibfnamefont {V.}~\bibnamefont {Cros}},\ }\href {\doibase
  10.1063/1.4862326} {\bibfield  {journal} {\bibinfo  {journal} {Applied
  Physics Letters}\ }\textbf {\bibinfo {volume} {104}},\ \bibinfo {pages}
  {022408} (\bibinfo {year} {2014})}\BibitemShut {NoStop}%
\bibitem [{\citenamefont {Van~Waeyenberge}\ \emph {et~al.}(2006)\citenamefont
  {Van~Waeyenberge}, \citenamefont {Puzic}, \citenamefont {Stoll},
  \citenamefont {Chou}, \citenamefont {Tyliszczak}, \citenamefont {Hertel},
  \citenamefont {F{\"a}hnle}, \citenamefont {Br{\"u}ckl}, \citenamefont {Rott},
  \citenamefont {Reiss}, \citenamefont {Neudecker}, \citenamefont {Weiss},
  \citenamefont {Back},\ and\ \citenamefont
  {Sch{\"u}tz}}]{van_waeyenberge_magnetic_2006}%
  \BibitemOpen
  \bibfield  {author} {\bibinfo {author} {\bibfnamefont {B.}~\bibnamefont
  {Van~Waeyenberge}}, \bibinfo {author} {\bibfnamefont {A.}~\bibnamefont
  {Puzic}}, \bibinfo {author} {\bibfnamefont {H.}~\bibnamefont {Stoll}},
  \bibinfo {author} {\bibfnamefont {K.~W.}\ \bibnamefont {Chou}}, \bibinfo
  {author} {\bibfnamefont {T.}~\bibnamefont {Tyliszczak}}, \bibinfo {author}
  {\bibfnamefont {R.}~\bibnamefont {Hertel}}, \bibinfo {author} {\bibfnamefont
  {M.}~\bibnamefont {F{\"a}hnle}}, \bibinfo {author} {\bibfnamefont
  {H.}~\bibnamefont {Br{\"u}ckl}}, \bibinfo {author} {\bibfnamefont
  {K.}~\bibnamefont {Rott}}, \bibinfo {author} {\bibfnamefont {G.}~\bibnamefont
  {Reiss}}, \bibinfo {author} {\bibfnamefont {I.}~\bibnamefont {Neudecker}},
  \bibinfo {author} {\bibfnamefont {D.}~\bibnamefont {Weiss}}, \bibinfo
  {author} {\bibfnamefont {C.~H.}\ \bibnamefont {Back}}, \ and\ \bibinfo
  {author} {\bibfnamefont {G.}~\bibnamefont {Sch{\"u}tz}},\ }\href {\doibase
  10.1038/nature05240} {\bibfield  {journal} {\bibinfo  {journal} {Nature}\
  }\textbf {\bibinfo {volume} {444}},\ \bibinfo {pages} {461} (\bibinfo {year}
  {2006})}\BibitemShut {NoStop}%
\bibitem [{\citenamefont {Petit-Watelot}\ \emph
  {et~al.}(2012{\natexlab{a}})\citenamefont {Petit-Watelot}, \citenamefont
  {Kim}, \citenamefont {Ruotolo}, \citenamefont {Otxoa}, \citenamefont
  {Bouzehouane}, \citenamefont {Grollier}, \citenamefont {Vansteenkiste},
  \citenamefont {Van~de Wiele}, \citenamefont {Cros},\ and\ \citenamefont
  {Devolder}}]{petit-watelot_commensurability_2012}%
  \BibitemOpen
  \bibfield  {author} {\bibinfo {author} {\bibfnamefont {S.}~\bibnamefont
  {Petit-Watelot}}, \bibinfo {author} {\bibfnamefont {J.-V.}\ \bibnamefont
  {Kim}}, \bibinfo {author} {\bibfnamefont {A.}~\bibnamefont {Ruotolo}},
  \bibinfo {author} {\bibfnamefont {R.~M.}\ \bibnamefont {Otxoa}}, \bibinfo
  {author} {\bibfnamefont {K.}~\bibnamefont {Bouzehouane}}, \bibinfo {author}
  {\bibfnamefont {J.}~\bibnamefont {Grollier}}, \bibinfo {author}
  {\bibfnamefont {A.}~\bibnamefont {Vansteenkiste}}, \bibinfo {author}
  {\bibfnamefont {B.}~\bibnamefont {Van~de Wiele}}, \bibinfo {author}
  {\bibfnamefont {V.}~\bibnamefont {Cros}}, \ and\ \bibinfo {author}
  {\bibfnamefont {T.}~\bibnamefont {Devolder}},\ }\href
  {http://dx.doi.org/10.1038/nphys2362} {\bibfield  {journal} {\bibinfo
  {journal} {Nature Physics}\ }\textbf {\bibinfo {volume} {8}},\ \bibinfo
  {pages} {682} (\bibinfo {year} {2012}{\natexlab{a}})}\BibitemShut {NoStop}%
\bibitem [{\citenamefont {Pylypovskyi}\ \emph {et~al.}(2013)\citenamefont
  {Pylypovskyi}, \citenamefont {Sheka}, \citenamefont {Kravchuk}, \citenamefont
  {Mertens},\ and\ \citenamefont {Gaididei}}]{pylypovskyi_regular_2013}%
  \BibitemOpen
  \bibfield  {author} {\bibinfo {author} {\bibfnamefont {O.~V.}\ \bibnamefont
  {Pylypovskyi}}, \bibinfo {author} {\bibfnamefont {D.~D.}\ \bibnamefont
  {Sheka}}, \bibinfo {author} {\bibfnamefont {V.~P.}\ \bibnamefont {Kravchuk}},
  \bibinfo {author} {\bibfnamefont {F.~G.}\ \bibnamefont {Mertens}}, \ and\
  \bibinfo {author} {\bibfnamefont {Y.}~\bibnamefont {Gaididei}},\ }\href
  {\doibase 10.1103/PhysRevB.88.014432} {\bibfield  {journal} {\bibinfo
  {journal} {Physical Review B}\ }\textbf {\bibinfo {volume} {88}},\ \bibinfo
  {pages} {014432} (\bibinfo {year} {2013})}\BibitemShut {NoStop}%
\bibitem [{\citenamefont {Williame}\ \emph {et~al.}(2019)\citenamefont
  {Williame}, \citenamefont {Difini~Accioly}, \citenamefont {Rontani},
  \citenamefont {Sciamanna},\ and\ \citenamefont
  {Kim}}]{williame_chaotic_2019}%
  \BibitemOpen
  \bibfield  {author} {\bibinfo {author} {\bibfnamefont {J.}~\bibnamefont
  {Williame}}, \bibinfo {author} {\bibfnamefont {A.}~\bibnamefont
  {Difini~Accioly}}, \bibinfo {author} {\bibfnamefont {D.}~\bibnamefont
  {Rontani}}, \bibinfo {author} {\bibfnamefont {M.}~\bibnamefont {Sciamanna}},
  \ and\ \bibinfo {author} {\bibfnamefont {J.-V.}\ \bibnamefont {Kim}},\ }\href
  {\doibase 10.1063/1.5095630} {\bibfield  {journal} {\bibinfo  {journal}
  {Applied Physics Letters}\ }\textbf {\bibinfo {volume} {114}},\ \bibinfo
  {pages} {232405} (\bibinfo {year} {2019})}\BibitemShut {NoStop}%
\bibitem [{\citenamefont {Bondarenko}\ \emph {et~al.}(2019)\citenamefont
  {Bondarenko}, \citenamefont {Holmgren}, \citenamefont {Li}, \citenamefont
  {Ivanov},\ and\ \citenamefont {Korenivski}}]{bondarenko_chaotic_2019}%
  \BibitemOpen
  \bibfield  {author} {\bibinfo {author} {\bibfnamefont {A.~V.}\ \bibnamefont
  {Bondarenko}}, \bibinfo {author} {\bibfnamefont {E.}~\bibnamefont
  {Holmgren}}, \bibinfo {author} {\bibfnamefont {Z.~W.}\ \bibnamefont {Li}},
  \bibinfo {author} {\bibfnamefont {B.~A.}\ \bibnamefont {Ivanov}}, \ and\
  \bibinfo {author} {\bibfnamefont {V.}~\bibnamefont {Korenivski}},\ }\href
  {\doibase 10.1103/PhysRevB.99.054402} {\bibfield  {journal} {\bibinfo
  {journal} {Physical Review B}\ }\textbf {\bibinfo {volume} {99}},\ \bibinfo
  {pages} {054402} (\bibinfo {year} {2019})}\BibitemShut {NoStop}%
\bibitem [{\citenamefont {Taniguchi}(2019)}]{taniguchi_synchronized_2019}%
  \BibitemOpen
  \bibfield  {author} {\bibinfo {author} {\bibfnamefont {T.}~\bibnamefont
  {Taniguchi}},\ }\href {\doibase 10.1016/j.jmmm.2019.03.090} {\bibfield
  {journal} {\bibinfo  {journal} {Journal of Magnetism and Magnetic Materials}\
  }\textbf {\bibinfo {volume} {483}},\ \bibinfo {pages} {281} (\bibinfo {year}
  {2019})}\BibitemShut {NoStop}%
\bibitem [{\citenamefont {Matsumoto}\ \emph {et~al.}(2019)\citenamefont
  {Matsumoto}, \citenamefont {Lequeux}, \citenamefont {Imamura},\ and\
  \citenamefont {Grollier}}]{matsumoto_chaos_2019}%
  \BibitemOpen
  \bibfield  {author} {\bibinfo {author} {\bibfnamefont {R.}~\bibnamefont
  {Matsumoto}}, \bibinfo {author} {\bibfnamefont {S.}~\bibnamefont {Lequeux}},
  \bibinfo {author} {\bibfnamefont {H.}~\bibnamefont {Imamura}}, \ and\
  \bibinfo {author} {\bibfnamefont {J.}~\bibnamefont {Grollier}},\ }\href
  {\doibase 10.1103/PhysRevApplied.11.044093} {\bibfield  {journal} {\bibinfo
  {journal} {Physical Review Applied}\ }\textbf {\bibinfo {volume} {11}},\
  \bibinfo {pages} {044093} (\bibinfo {year} {2019})}\BibitemShut {NoStop}%
\bibitem [{\citenamefont {Sciamanna}\ and\ \citenamefont
  {Shore}(2015)}]{sciamanna_physics_2015}%
  \BibitemOpen
  \bibfield  {author} {\bibinfo {author} {\bibfnamefont {M.}~\bibnamefont
  {Sciamanna}}\ and\ \bibinfo {author} {\bibfnamefont {K.~A.}\ \bibnamefont
  {Shore}},\ }\href {\doibase 10.1038/nphoton.2014.326} {\bibfield  {journal}
  {\bibinfo  {journal} {Nature Photonics}\ }\textbf {\bibinfo {volume} {9}},\
  \bibinfo {pages} {151} (\bibinfo {year} {2015})}\BibitemShut {NoStop}%
\bibitem [{\citenamefont {Martin}\ \emph {et~al.}(2011)\citenamefont {Martin},
  \citenamefont {de~Mestier}, \citenamefont {Thirion}, \citenamefont {Hoarau},
  \citenamefont {Conraux}, \citenamefont {Baraduc},\ and\ \citenamefont
  {Di{\'e}ny}}]{martin_parametric_2011}%
  \BibitemOpen
  \bibfield  {author} {\bibinfo {author} {\bibfnamefont {S.~Y.}\ \bibnamefont
  {Martin}}, \bibinfo {author} {\bibfnamefont {N.}~\bibnamefont {de~Mestier}},
  \bibinfo {author} {\bibfnamefont {C.}~\bibnamefont {Thirion}}, \bibinfo
  {author} {\bibfnamefont {C.}~\bibnamefont {Hoarau}}, \bibinfo {author}
  {\bibfnamefont {Y.}~\bibnamefont {Conraux}}, \bibinfo {author} {\bibfnamefont
  {C.}~\bibnamefont {Baraduc}}, \ and\ \bibinfo {author} {\bibfnamefont
  {B.}~\bibnamefont {Di{\'e}ny}},\ }\href {\doibase 10.1103/PhysRevB.84.144434}
  {\bibfield  {journal} {\bibinfo  {journal} {Physical Review B}\ }\textbf
  {\bibinfo {volume} {84}},\ \bibinfo {pages} {144434} (\bibinfo {year}
  {2011})}\BibitemShut {NoStop}%
\bibitem [{\citenamefont {Manfrini}\ \emph {et~al.}(2011)\citenamefont
  {Manfrini}, \citenamefont {Devolder}, \citenamefont {Kim}, \citenamefont
  {Crozat}, \citenamefont {Chappert}, \citenamefont {Van~Roy},\ and\
  \citenamefont {Lagae}}]{manfrini_frequency_2011}%
  \BibitemOpen
  \bibfield  {author} {\bibinfo {author} {\bibfnamefont {M.}~\bibnamefont
  {Manfrini}}, \bibinfo {author} {\bibfnamefont {T.}~\bibnamefont {Devolder}},
  \bibinfo {author} {\bibfnamefont {J.-V.}\ \bibnamefont {Kim}}, \bibinfo
  {author} {\bibfnamefont {P.}~\bibnamefont {Crozat}}, \bibinfo {author}
  {\bibfnamefont {C.}~\bibnamefont {Chappert}}, \bibinfo {author}
  {\bibfnamefont {W.}~\bibnamefont {Van~Roy}}, \ and\ \bibinfo {author}
  {\bibfnamefont {L.}~\bibnamefont {Lagae}},\ }\href {\doibase
  10.1063/1.3581099} {\bibfield  {journal} {\bibinfo  {journal} {Journal of
  Applied Physics}\ }\textbf {\bibinfo {volume} {109}},\ \bibinfo {pages}
  {083940} (\bibinfo {year} {2011})}\BibitemShut {NoStop}%
\bibitem [{\citenamefont {Ivanov}\ and\ \citenamefont
  {Zaspel}(2007)}]{ivanov_excitation_2007}%
  \BibitemOpen
  \bibfield  {author} {\bibinfo {author} {\bibfnamefont {B.~A.}\ \bibnamefont
  {Ivanov}}\ and\ \bibinfo {author} {\bibfnamefont {C.~E.}\ \bibnamefont
  {Zaspel}},\ }\href {\doibase 10.1103/PhysRevLett.99.247208} {\bibfield
  {journal} {\bibinfo  {journal} {Physical Review Letters}\ }\textbf {\bibinfo
  {volume} {99}},\ \bibinfo {pages} {247208} (\bibinfo {year}
  {2007})}\BibitemShut {NoStop}%
\bibitem [{\citenamefont {Zhang}\ and\ \citenamefont
  {Li}(2004)}]{zhang_roles_2004}%
  \BibitemOpen
  \bibfield  {author} {\bibinfo {author} {\bibfnamefont {S.}~\bibnamefont
  {Zhang}}\ and\ \bibinfo {author} {\bibfnamefont {Z.}~\bibnamefont {Li}},\
  }\href {\doibase 10.1103/PhysRevLett.93.127204} {\bibfield  {journal}
  {\bibinfo  {journal} {Physical Review Letters}\ }\textbf {\bibinfo {volume}
  {93}},\ \bibinfo {pages} {127204} (\bibinfo {year} {2004})}\BibitemShut
  {NoStop}%
\bibitem [{\citenamefont {Bouzehouane}\ \emph {et~al.}(2003)\citenamefont
  {Bouzehouane}, \citenamefont {Fusil}, \citenamefont {Bibes}, \citenamefont
  {Carrey}, \citenamefont {Blon}, \citenamefont {Le~D{\^u}}, \citenamefont
  {Seneor}, \citenamefont {Cros},\ and\ \citenamefont
  {Vila}}]{bouzehouane_nanolithography_2003}%
  \BibitemOpen
  \bibfield  {author} {\bibinfo {author} {\bibfnamefont {K.}~\bibnamefont
  {Bouzehouane}}, \bibinfo {author} {\bibfnamefont {S.}~\bibnamefont {Fusil}},
  \bibinfo {author} {\bibfnamefont {M.}~\bibnamefont {Bibes}}, \bibinfo
  {author} {\bibfnamefont {J.}~\bibnamefont {Carrey}}, \bibinfo {author}
  {\bibfnamefont {T.}~\bibnamefont {Blon}}, \bibinfo {author} {\bibfnamefont
  {M.}~\bibnamefont {Le~D{\^u}}}, \bibinfo {author} {\bibfnamefont
  {P.}~\bibnamefont {Seneor}}, \bibinfo {author} {\bibfnamefont
  {V.}~\bibnamefont {Cros}}, \ and\ \bibinfo {author} {\bibfnamefont
  {L.}~\bibnamefont {Vila}},\ }\href {\doibase 10.1021/nl034610j} {\bibfield
  {journal} {\bibinfo  {journal} {Nano Letters}\ }\textbf {\bibinfo {volume}
  {3}},\ \bibinfo {pages} {1599} (\bibinfo {year} {2003})}\BibitemShut
  {NoStop}%
\bibitem [{\citenamefont {Ruotolo}\ \emph {et~al.}(2009)\citenamefont
  {Ruotolo}, \citenamefont {Cros}, \citenamefont {Georges}, \citenamefont
  {Dussaux}, \citenamefont {Grollier}, \citenamefont {Deranlot}, \citenamefont
  {Guillemet}, \citenamefont {Bouzehouane}, \citenamefont {Fusil},\ and\
  \citenamefont {Fert}}]{ruotolo_phase-locking_2009}%
  \BibitemOpen
  \bibfield  {author} {\bibinfo {author} {\bibfnamefont {A.}~\bibnamefont
  {Ruotolo}}, \bibinfo {author} {\bibfnamefont {V.}~\bibnamefont {Cros}},
  \bibinfo {author} {\bibfnamefont {B.}~\bibnamefont {Georges}}, \bibinfo
  {author} {\bibfnamefont {A.}~\bibnamefont {Dussaux}}, \bibinfo {author}
  {\bibfnamefont {J.}~\bibnamefont {Grollier}}, \bibinfo {author}
  {\bibfnamefont {C.}~\bibnamefont {Deranlot}}, \bibinfo {author}
  {\bibfnamefont {R.}~\bibnamefont {Guillemet}}, \bibinfo {author}
  {\bibfnamefont {K.}~\bibnamefont {Bouzehouane}}, \bibinfo {author}
  {\bibfnamefont {S.}~\bibnamefont {Fusil}}, \ and\ \bibinfo {author}
  {\bibfnamefont {A.}~\bibnamefont {Fert}},\ }\href {\doibase
  10.1038/nnano.2009.143} {\bibfield  {journal} {\bibinfo  {journal} {Nature
  Nanotechnology}\ }\textbf {\bibinfo {volume} {4}},\ \bibinfo {pages} {528}
  (\bibinfo {year} {2009})}\BibitemShut {NoStop}%
\bibitem [{\citenamefont {Jaromirska}\ \emph {et~al.}(2011)\citenamefont
  {Jaromirska}, \citenamefont {Lopez-Diaz}, \citenamefont {Ruotolo},
  \citenamefont {Grollier}, \citenamefont {Cros},\ and\ \citenamefont
  {Berkov}}]{jaromirska_influence_2011}%
  \BibitemOpen
  \bibfield  {author} {\bibinfo {author} {\bibfnamefont {E.}~\bibnamefont
  {Jaromirska}}, \bibinfo {author} {\bibfnamefont {L.}~\bibnamefont
  {Lopez-Diaz}}, \bibinfo {author} {\bibfnamefont {A.}~\bibnamefont {Ruotolo}},
  \bibinfo {author} {\bibfnamefont {J.}~\bibnamefont {Grollier}}, \bibinfo
  {author} {\bibfnamefont {V.}~\bibnamefont {Cros}}, \ and\ \bibinfo {author}
  {\bibfnamefont {D.}~\bibnamefont {Berkov}},\ }\href
  {https://link.aps.org/doi/10.1103/PhysRevB.83.094419} {\bibfield  {journal}
  {\bibinfo  {journal} {Physical Review B}\ }\textbf {\bibinfo {volume} {83}}
  (\bibinfo {year} {2011})}\BibitemShut {NoStop}%
\bibitem [{\citenamefont {Petit-Watelot}\ \emph
  {et~al.}(2012{\natexlab{b}})\citenamefont {Petit-Watelot}, \citenamefont
  {Otxoa}, \citenamefont {Manfrini}, \citenamefont {Van~Roy}, \citenamefont
  {Lagae}, \citenamefont {Kim},\ and\ \citenamefont
  {Devolder}}]{petit-watelot_understanding_2012}%
  \BibitemOpen
  \bibfield  {author} {\bibinfo {author} {\bibfnamefont {S.}~\bibnamefont
  {Petit-Watelot}}, \bibinfo {author} {\bibfnamefont {R.~M.}\ \bibnamefont
  {Otxoa}}, \bibinfo {author} {\bibfnamefont {M.}~\bibnamefont {Manfrini}},
  \bibinfo {author} {\bibfnamefont {W.}~\bibnamefont {Van~Roy}}, \bibinfo
  {author} {\bibfnamefont {L.}~\bibnamefont {Lagae}}, \bibinfo {author}
  {\bibfnamefont {J.-V.}\ \bibnamefont {Kim}}, \ and\ \bibinfo {author}
  {\bibfnamefont {T.}~\bibnamefont {Devolder}},\ }\href {\doibase
  10.1103/PhysRevLett.109.267205} {\bibfield  {journal} {\bibinfo  {journal}
  {Physical Review Letters}\ }\textbf {\bibinfo {volume} {109}},\ \bibinfo
  {pages} {267205} (\bibinfo {year} {2012}{\natexlab{b}})}\BibitemShut
  {NoStop}%
\bibitem [{\citenamefont {Petit-Watelot}\ \emph
  {et~al.}(2012{\natexlab{c}})\citenamefont {Petit-Watelot}, \citenamefont
  {Otxoa},\ and\ \citenamefont {Manfrini}}]{petit-watelot_electrical_2012}%
  \BibitemOpen
  \bibfield  {author} {\bibinfo {author} {\bibfnamefont {S.}~\bibnamefont
  {Petit-Watelot}}, \bibinfo {author} {\bibfnamefont {R.~M.}\ \bibnamefont
  {Otxoa}}, \ and\ \bibinfo {author} {\bibfnamefont {M.}~\bibnamefont
  {Manfrini}},\ }\href {\doibase 10.1063/1.3687915} {\bibfield  {journal}
  {\bibinfo  {journal} {Applied Physics Letters}\ }\textbf {\bibinfo {volume}
  {100}},\ \bibinfo {pages} {083507} (\bibinfo {year}
  {2012}{\natexlab{c}})}\BibitemShut {NoStop}%
\bibitem [{\citenamefont {Vansteenkiste}\ \emph {et~al.}(2014)\citenamefont
  {Vansteenkiste}, \citenamefont {Leliaert}, \citenamefont {Dvornik},
  \citenamefont {Helsen}, \citenamefont {Garcia-Sanchez},\ and\ \citenamefont
  {Waeyenberge}}]{vansteenkiste_design_2014}%
  \BibitemOpen
  \bibfield  {author} {\bibinfo {author} {\bibfnamefont {A.}~\bibnamefont
  {Vansteenkiste}}, \bibinfo {author} {\bibfnamefont {J.}~\bibnamefont
  {Leliaert}}, \bibinfo {author} {\bibfnamefont {M.}~\bibnamefont {Dvornik}},
  \bibinfo {author} {\bibfnamefont {M.}~\bibnamefont {Helsen}}, \bibinfo
  {author} {\bibfnamefont {F.}~\bibnamefont {Garcia-Sanchez}}, \ and\ \bibinfo
  {author} {\bibfnamefont {B.~V.}\ \bibnamefont {Waeyenberge}},\ }\href
  {\doibase 10.1063/1.4899186} {\bibfield  {journal} {\bibinfo  {journal} {AIP
  Advances}\ }\textbf {\bibinfo {volume} {4}},\ \bibinfo {pages} {107133}
  (\bibinfo {year} {2014})}\BibitemShut {NoStop}%
\bibitem [{\citenamefont {Landau}\ and\ \citenamefont
  {Lifshitz}(1935)}]{landau_theory_1935}%
  \BibitemOpen
  \bibfield  {author} {\bibinfo {author} {\bibfnamefont {L.}~\bibnamefont
  {Landau}}\ and\ \bibinfo {author} {\bibfnamefont {E.}~\bibnamefont
  {Lifshitz}},\ }\href
  {http://ujp.bitp.kiev.ua/files/journals/53/si/53SI06p.pdf} {\bibfield
  {journal} {\bibinfo  {journal} {Physikalische Zeitschrift der Sowjetunion}\
  }\textbf {\bibinfo {volume} {8}},\ \bibinfo {pages} {153} (\bibinfo {year}
  {1935})}\BibitemShut {NoStop}%
\bibitem [{\citenamefont {Gilbert}(1955)}]{gilbert_lagrangian_1955}%
  \BibitemOpen
  \bibfield  {author} {\bibinfo {author} {\bibfnamefont {T.~L.}\ \bibnamefont
  {Gilbert}},\ }\href@noop {} {\bibfield  {journal} {\bibinfo  {journal}
  {Physical Review}\ }\textbf {\bibinfo {volume} {100}},\ \bibinfo {pages}
  {1235} (\bibinfo {year} {1955})}\BibitemShut {NoStop}%
\bibitem [{\citenamefont {Gilbert}(2004)}]{gilbert_phenomenological_2004}%
  \BibitemOpen
  \bibfield  {author} {\bibinfo {author} {\bibfnamefont {T.~L.}\ \bibnamefont
  {Gilbert}},\ }\href {\doibase 10.1109/TMAG.2004.836740} {\bibfield  {journal}
  {\bibinfo  {journal} {IEEE Transactions on Magnetics}\ }\textbf {\bibinfo
  {volume} {40}},\ \bibinfo {pages} {3443} (\bibinfo {year}
  {2004})}\BibitemShut {NoStop}%
\bibitem [{\citenamefont {Devolder}\ \emph {et~al.}(2011)\citenamefont
  {Devolder}, \citenamefont {Kim}, \citenamefont {Petit-Watelot}, \citenamefont
  {Otxoa}, \citenamefont {Chappert}, \citenamefont {Manfrini}, \citenamefont
  {Van~Roy},\ and\ \citenamefont {Lagae}}]{devolder_vortex_2011}%
  \BibitemOpen
  \bibfield  {author} {\bibinfo {author} {\bibfnamefont {T.}~\bibnamefont
  {Devolder}}, \bibinfo {author} {\bibfnamefont {J.-V.}\ \bibnamefont {Kim}},
  \bibinfo {author} {\bibfnamefont {S.}~\bibnamefont {Petit-Watelot}}, \bibinfo
  {author} {\bibfnamefont {R.}~\bibnamefont {Otxoa}}, \bibinfo {author}
  {\bibfnamefont {C.}~\bibnamefont {Chappert}}, \bibinfo {author}
  {\bibfnamefont {M.}~\bibnamefont {Manfrini}}, \bibinfo {author}
  {\bibfnamefont {W.}~\bibnamefont {Van~Roy}}, \ and\ \bibinfo {author}
  {\bibfnamefont {L.}~\bibnamefont {Lagae}},\ }\href {\doibase
  10.1109/TMAG.2010.2101056} {\bibfield  {journal} {\bibinfo  {journal} {IEEE
  Transactions on Magnetics}\ }\textbf {\bibinfo {volume} {47}},\ \bibinfo
  {pages} {1595} (\bibinfo {year} {2011})}\BibitemShut {NoStop}%
\bibitem [{\citenamefont {Nakatani}\ \emph {et~al.}(2008)\citenamefont
  {Nakatani}, \citenamefont {Shibata}, \citenamefont {Tatara}, \citenamefont
  {Kohno}, \citenamefont {Thiaville},\ and\ \citenamefont
  {Miltat}}]{nakatani_nucleation_2008}%
  \BibitemOpen
  \bibfield  {author} {\bibinfo {author} {\bibfnamefont {Y.}~\bibnamefont
  {Nakatani}}, \bibinfo {author} {\bibfnamefont {J.}~\bibnamefont {Shibata}},
  \bibinfo {author} {\bibfnamefont {G.}~\bibnamefont {Tatara}}, \bibinfo
  {author} {\bibfnamefont {H.}~\bibnamefont {Kohno}}, \bibinfo {author}
  {\bibfnamefont {A.}~\bibnamefont {Thiaville}}, \ and\ \bibinfo {author}
  {\bibfnamefont {J.}~\bibnamefont {Miltat}},\ }\href {\doibase
  10.1103/PhysRevB.77.014439} {\bibfield  {journal} {\bibinfo  {journal}
  {Physical Review B}\ }\textbf {\bibinfo {volume} {77}},\ \bibinfo {pages}
  {014439} (\bibinfo {year} {2008})}\BibitemShut {NoStop}%
\bibitem [{\citenamefont {Otxoa}\ \emph {et~al.}(2015)\citenamefont {Otxoa},
  \citenamefont {Petit-Watelot}, \citenamefont {Manfrini}, \citenamefont
  {Radu}, \citenamefont {Thean}, \citenamefont {Kim},\ and\ \citenamefont
  {Devolder}}]{otxoa_dynamical_2015}%
  \BibitemOpen
  \bibfield  {author} {\bibinfo {author} {\bibfnamefont {R.}~\bibnamefont
  {Otxoa}}, \bibinfo {author} {\bibfnamefont {S.}~\bibnamefont
  {Petit-Watelot}}, \bibinfo {author} {\bibfnamefont {M.}~\bibnamefont
  {Manfrini}}, \bibinfo {author} {\bibfnamefont {I.}~\bibnamefont {Radu}},
  \bibinfo {author} {\bibfnamefont {A.}~\bibnamefont {Thean}}, \bibinfo
  {author} {\bibfnamefont {J.-V.}\ \bibnamefont {Kim}}, \ and\ \bibinfo
  {author} {\bibfnamefont {T.}~\bibnamefont {Devolder}},\ }\href {\doibase
  10.1016/j.jmmm.2015.06.057} {\bibfield  {journal} {\bibinfo  {journal}
  {Journal of Magnetism and Magnetic Materials}\ }\textbf {\bibinfo {volume}
  {394}},\ \bibinfo {pages} {292} (\bibinfo {year} {2015})}\BibitemShut
  {NoStop}%
\bibitem [{\citenamefont {Devolder}\ \emph {et~al.}(2019)\citenamefont
  {Devolder}, \citenamefont {Rontani}, \citenamefont {Petit-Watelot},
  \citenamefont {Bouzehouane}, \citenamefont {Andrieu}, \citenamefont
  {L{\'e}tang}, \citenamefont {Yoo}, \citenamefont {Adam}, \citenamefont
  {Chappert}, \citenamefont {Girod}, \citenamefont {Cros}, \citenamefont
  {Sciamanna},\ and\ \citenamefont {Kim}}]{devolder_chaos_2019}%
  \BibitemOpen
  \bibfield  {author} {\bibinfo {author} {\bibfnamefont {T.}~\bibnamefont
  {Devolder}}, \bibinfo {author} {\bibfnamefont {D.}~\bibnamefont {Rontani}},
  \bibinfo {author} {\bibfnamefont {S.}~\bibnamefont {Petit-Watelot}}, \bibinfo
  {author} {\bibfnamefont {K.}~\bibnamefont {Bouzehouane}}, \bibinfo {author}
  {\bibfnamefont {S.}~\bibnamefont {Andrieu}}, \bibinfo {author} {\bibfnamefont
  {J.}~\bibnamefont {L{\'e}tang}}, \bibinfo {author} {\bibfnamefont {M.-W.}\
  \bibnamefont {Yoo}}, \bibinfo {author} {\bibfnamefont {J.-P.}\ \bibnamefont
  {Adam}}, \bibinfo {author} {\bibfnamefont {C.}~\bibnamefont {Chappert}},
  \bibinfo {author} {\bibfnamefont {S.}~\bibnamefont {Girod}}, \bibinfo
  {author} {\bibfnamefont {V.}~\bibnamefont {Cros}}, \bibinfo {author}
  {\bibfnamefont {M.}~\bibnamefont {Sciamanna}}, \ and\ \bibinfo {author}
  {\bibfnamefont {J.-V.}\ \bibnamefont {Kim}},\ }\href {\doibase
  10.1103/PhysRevLett.123.147701} {\bibfield  {journal} {\bibinfo  {journal}
  {Physical Review Letters}\ }\textbf {\bibinfo {volume} {123}},\ \bibinfo
  {pages} {147701} (\bibinfo {year} {2019})}\BibitemShut {NoStop}%
\bibitem [{\citenamefont {Kuepferling}\ \emph {et~al.}(2010)\citenamefont
  {Kuepferling}, \citenamefont {Serpico}, \citenamefont {Pufall}, \citenamefont
  {Rippard}, \citenamefont {Wallis}, \citenamefont {Imtiaz}, \citenamefont
  {Krivosik}, \citenamefont {Pasquale},\ and\ \citenamefont
  {Kabos}}]{kuepferling_two_2010}%
  \BibitemOpen
  \bibfield  {author} {\bibinfo {author} {\bibfnamefont {M.}~\bibnamefont
  {Kuepferling}}, \bibinfo {author} {\bibfnamefont {C.}~\bibnamefont
  {Serpico}}, \bibinfo {author} {\bibfnamefont {M.}~\bibnamefont {Pufall}},
  \bibinfo {author} {\bibfnamefont {W.}~\bibnamefont {Rippard}}, \bibinfo
  {author} {\bibfnamefont {T.~M.}\ \bibnamefont {Wallis}}, \bibinfo {author}
  {\bibfnamefont {A.}~\bibnamefont {Imtiaz}}, \bibinfo {author} {\bibfnamefont
  {P.}~\bibnamefont {Krivosik}}, \bibinfo {author} {\bibfnamefont
  {M.}~\bibnamefont {Pasquale}}, \ and\ \bibinfo {author} {\bibfnamefont
  {P.}~\bibnamefont {Kabos}},\ }\href {\doibase 10.1063/1.3455883} {\bibfield
  {journal} {\bibinfo  {journal} {Applied Physics Letters}\ }\textbf {\bibinfo
  {volume} {96}},\ \bibinfo {pages} {252507} (\bibinfo {year}
  {2010})}\BibitemShut {NoStop}%
\bibitem [{\citenamefont {Wang}\ \emph {et~al.}(2011)\citenamefont {Wang},
  \citenamefont {Wang}, \citenamefont {Qin}, \citenamefont {Yeung},
  \citenamefont {Kwok}, \citenamefont {Wong}, \citenamefont {Xue},
  \citenamefont {Chu}, \citenamefont {Leung},\ and\ \citenamefont
  {Ruotolo}}]{wang_multiple-mode_2011}%
  \BibitemOpen
  \bibfield  {author} {\bibinfo {author} {\bibfnamefont {N.}~\bibnamefont
  {Wang}}, \bibinfo {author} {\bibfnamefont {X.~L.}\ \bibnamefont {Wang}},
  \bibinfo {author} {\bibfnamefont {W.}~\bibnamefont {Qin}}, \bibinfo {author}
  {\bibfnamefont {S.~H.}\ \bibnamefont {Yeung}}, \bibinfo {author}
  {\bibfnamefont {D.~T.~K.}\ \bibnamefont {Kwok}}, \bibinfo {author}
  {\bibfnamefont {H.~F.}\ \bibnamefont {Wong}}, \bibinfo {author}
  {\bibfnamefont {Q.}~\bibnamefont {Xue}}, \bibinfo {author} {\bibfnamefont
  {P.~K.}\ \bibnamefont {Chu}}, \bibinfo {author} {\bibfnamefont {C.~W.}\
  \bibnamefont {Leung}}, \ and\ \bibinfo {author} {\bibfnamefont
  {A.}~\bibnamefont {Ruotolo}},\ }\href {\doibase 10.1063/1.3600328} {\bibfield
   {journal} {\bibinfo  {journal} {Applied Physics Letters}\ }\textbf {\bibinfo
  {volume} {98}},\ \bibinfo {pages} {242506} (\bibinfo {year}
  {2011})}\BibitemShut {NoStop}%
\bibitem [{\citenamefont {Keatley}\ \emph {et~al.}(2016)\citenamefont
  {Keatley}, \citenamefont {Sani}, \citenamefont {Hrkac}, \citenamefont
  {Mohseni}, \citenamefont {D{\"u}rrenfeld}, \citenamefont {Loughran},
  \citenamefont {{\r A}kerman},\ and\ \citenamefont
  {Hicken}}]{keatley_direct_2016}%
  \BibitemOpen
  \bibfield  {author} {\bibinfo {author} {\bibfnamefont {P.~S.}\ \bibnamefont
  {Keatley}}, \bibinfo {author} {\bibfnamefont {S.~R.}\ \bibnamefont {Sani}},
  \bibinfo {author} {\bibfnamefont {G.}~\bibnamefont {Hrkac}}, \bibinfo
  {author} {\bibfnamefont {S.~M.}\ \bibnamefont {Mohseni}}, \bibinfo {author}
  {\bibfnamefont {P.}~\bibnamefont {D{\"u}rrenfeld}}, \bibinfo {author}
  {\bibfnamefont {T.~H.~J.}\ \bibnamefont {Loughran}}, \bibinfo {author}
  {\bibfnamefont {J.}~\bibnamefont {{\r A}kerman}}, \ and\ \bibinfo {author}
  {\bibfnamefont {R.~J.}\ \bibnamefont {Hicken}},\ }\href {\doibase
  10.1103/PhysRevB.94.060402} {\bibfield  {journal} {\bibinfo  {journal}
  {Physical Review B}\ }\textbf {\bibinfo {volume} {94}},\ \bibinfo {pages}
  {060402} (\bibinfo {year} {2016})}\BibitemShut {NoStop}%
\bibitem [{\citenamefont {Keatley}\ \emph {et~al.}(2017)\citenamefont
  {Keatley}, \citenamefont {Sani}, \citenamefont {Hrkac}, \citenamefont
  {Mohseni}, \citenamefont {D{\"u}rrenfeld}, \citenamefont {{\r A}kerman},\
  and\ \citenamefont {Hicken}}]{keatley_imaging_2017}%
  \BibitemOpen
  \bibfield  {author} {\bibinfo {author} {\bibfnamefont {P.~S.}\ \bibnamefont
  {Keatley}}, \bibinfo {author} {\bibfnamefont {S.~R.}\ \bibnamefont {Sani}},
  \bibinfo {author} {\bibfnamefont {G.}~\bibnamefont {Hrkac}}, \bibinfo
  {author} {\bibfnamefont {S.~M.}\ \bibnamefont {Mohseni}}, \bibinfo {author}
  {\bibfnamefont {P.}~\bibnamefont {D{\"u}rrenfeld}}, \bibinfo {author}
  {\bibfnamefont {J.}~\bibnamefont {{\r A}kerman}}, \ and\ \bibinfo {author}
  {\bibfnamefont {R.~J.}\ \bibnamefont {Hicken}},\ }\href {\doibase
  10.1088/1361-6463/aa628a} {\bibfield  {journal} {\bibinfo  {journal} {Journal
  of Physics D: Applied Physics}\ }\textbf {\bibinfo {volume} {50}},\ \bibinfo
  {pages} {164003} (\bibinfo {year} {2017})}\BibitemShut {NoStop}%
\bibitem [{\citenamefont {Yamada}\ \emph {et~al.}(2007)\citenamefont {Yamada},
  \citenamefont {Kasai}, \citenamefont {Nakatani}, \citenamefont {Kobayashi},
  \citenamefont {Kohno}, \citenamefont {Thiaville},\ and\ \citenamefont
  {Ono}}]{yamada_electrical_2007}%
  \BibitemOpen
  \bibfield  {author} {\bibinfo {author} {\bibfnamefont {K.}~\bibnamefont
  {Yamada}}, \bibinfo {author} {\bibfnamefont {S.}~\bibnamefont {Kasai}},
  \bibinfo {author} {\bibfnamefont {Y.}~\bibnamefont {Nakatani}}, \bibinfo
  {author} {\bibfnamefont {K.}~\bibnamefont {Kobayashi}}, \bibinfo {author}
  {\bibfnamefont {H.}~\bibnamefont {Kohno}}, \bibinfo {author} {\bibfnamefont
  {A.}~\bibnamefont {Thiaville}}, \ and\ \bibinfo {author} {\bibfnamefont
  {T.}~\bibnamefont {Ono}},\ }\href {\doibase 10.1038/nmat1867} {\bibfield
  {journal} {\bibinfo  {journal} {Nature Materials}\ }\textbf {\bibinfo
  {volume} {6}},\ \bibinfo {pages} {270} (\bibinfo {year} {2007})}\BibitemShut
  {NoStop}%
\bibitem [{\citenamefont {Guslienko}\ \emph {et~al.}(2008)\citenamefont
  {Guslienko}, \citenamefont {Lee},\ and\ \citenamefont
  {Kim}}]{guslienko_dynamic_2008}%
  \BibitemOpen
  \bibfield  {author} {\bibinfo {author} {\bibfnamefont {K.~Y.}\ \bibnamefont
  {Guslienko}}, \bibinfo {author} {\bibfnamefont {K.-S.}\ \bibnamefont {Lee}},
  \ and\ \bibinfo {author} {\bibfnamefont {S.-K.}\ \bibnamefont {Kim}},\ }\href
  {\doibase 10.1103/PhysRevLett.100.027203} {\bibfield  {journal} {\bibinfo
  {journal} {Physical Review Letters}\ }\textbf {\bibinfo {volume} {100}},\
  \bibinfo {pages} {027203} (\bibinfo {year} {2008})}\BibitemShut {NoStop}%
\bibitem [{\citenamefont {Hertel}\ and\ \citenamefont
  {Schneider}(2006)}]{hertel_exchange_2006}%
  \BibitemOpen
  \bibfield  {author} {\bibinfo {author} {\bibfnamefont {R.}~\bibnamefont
  {Hertel}}\ and\ \bibinfo {author} {\bibfnamefont {C.~M.}\ \bibnamefont
  {Schneider}},\ }\href {\doibase 10.1103/PhysRevLett.97.177202} {\bibfield
  {journal} {\bibinfo  {journal} {Physical Review Letters}\ }\textbf {\bibinfo
  {volume} {97}},\ \bibinfo {pages} {177202} (\bibinfo {year}
  {2006})}\BibitemShut {NoStop}%
\bibitem [{\citenamefont {Thiele}(1973)}]{thiele_steady-state_1973}%
  \BibitemOpen
  \bibfield  {author} {\bibinfo {author} {\bibfnamefont {A.~A.}\ \bibnamefont
  {Thiele}},\ }\href {\doibase 10.1103/PhysRevLett.30.230} {\bibfield
  {journal} {\bibinfo  {journal} {Physical Review Letters}\ }\textbf {\bibinfo
  {volume} {30}},\ \bibinfo {pages} {230} (\bibinfo {year} {1973})}\BibitemShut
  {NoStop}%
\bibitem [{\citenamefont {Huber}(1982)}]{huber_dynamics_1982}%
  \BibitemOpen
  \bibfield  {author} {\bibinfo {author} {\bibfnamefont {D.~L.}\ \bibnamefont
  {Huber}},\ }\href {\doibase 10.1103/PhysRevB.26.3758} {\bibfield  {journal}
  {\bibinfo  {journal} {Physical Review B}\ }\textbf {\bibinfo {volume} {26}},\
  \bibinfo {pages} {3758} (\bibinfo {year} {1982})}\BibitemShut {NoStop}%
\bibitem [{\citenamefont {Kim}(2012)}]{Kim:2012du}%
  \BibitemOpen
  \bibfield  {author} {\bibinfo {author} {\bibfnamefont {J.-V.}\ \bibnamefont
  {Kim}},\ }in\ \href@noop {} {\emph {\bibinfo {booktitle} {Solid State
  Physics}}},\ \bibinfo {editor} {edited by\ \bibinfo {editor} {\bibfnamefont
  {R.~E.}\ \bibnamefont {Camley}}\ and\ \bibinfo {editor} {\bibfnamefont
  {R.~L.}\ \bibnamefont {Stamps}}}\ (\bibinfo  {publisher} {Academic Press},\
  \bibinfo {year} {2012})\ pp.\ \bibinfo {pages} {217--294}\BibitemShut
  {NoStop}%
\bibitem [{\citenamefont {Urazhdin}\ \emph {et~al.}(2010)\citenamefont
  {Urazhdin}, \citenamefont {Tabor}, \citenamefont {Tiberkevich},\ and\
  \citenamefont {Slavin}}]{urazhdin_fractional_2010}%
  \BibitemOpen
  \bibfield  {author} {\bibinfo {author} {\bibfnamefont {S.}~\bibnamefont
  {Urazhdin}}, \bibinfo {author} {\bibfnamefont {P.}~\bibnamefont {Tabor}},
  \bibinfo {author} {\bibfnamefont {V.}~\bibnamefont {Tiberkevich}}, \ and\
  \bibinfo {author} {\bibfnamefont {A.}~\bibnamefont {Slavin}},\ }\href
  {\doibase 10.1103/PhysRevLett.105.104101} {\bibfield  {journal} {\bibinfo
  {journal} {Physical Review Letters}\ }\textbf {\bibinfo {volume} {105}},\
  \bibinfo {pages} {104101} (\bibinfo {year} {2010})}\BibitemShut {NoStop}%
\bibitem [{\citenamefont {Singh}\ \emph {et~al.}(2017)\citenamefont {Singh},
  \citenamefont {Konishi}, \citenamefont {Bhuktare}, \citenamefont {Bose},
  \citenamefont {Miwa}, \citenamefont {Fukushima}, \citenamefont {Yakushiji},
  \citenamefont {Yuasa}, \citenamefont {Kubota}, \citenamefont {Suzuki},\ and\
  \citenamefont {Tulapurkar}}]{singh_integer_2017}%
  \BibitemOpen
  \bibfield  {author} {\bibinfo {author} {\bibfnamefont {H.}~\bibnamefont
  {Singh}}, \bibinfo {author} {\bibfnamefont {K.}~\bibnamefont {Konishi}},
  \bibinfo {author} {\bibfnamefont {S.}~\bibnamefont {Bhuktare}}, \bibinfo
  {author} {\bibfnamefont {A.}~\bibnamefont {Bose}}, \bibinfo {author}
  {\bibfnamefont {S.}~\bibnamefont {Miwa}}, \bibinfo {author} {\bibfnamefont
  {A.}~\bibnamefont {Fukushima}}, \bibinfo {author} {\bibfnamefont
  {K.}~\bibnamefont {Yakushiji}}, \bibinfo {author} {\bibfnamefont
  {S.}~\bibnamefont {Yuasa}}, \bibinfo {author} {\bibfnamefont
  {H.}~\bibnamefont {Kubota}}, \bibinfo {author} {\bibfnamefont
  {Y.}~\bibnamefont {Suzuki}}, \ and\ \bibinfo {author} {\bibfnamefont {A.~A.}\
  \bibnamefont {Tulapurkar}},\ }\href {\doibase
  10.1103/PhysRevApplied.8.064011} {\bibfield  {journal} {\bibinfo  {journal}
  {Physical Review Applied}\ }\textbf {\bibinfo {volume} {8}},\ \bibinfo
  {pages} {064011} (\bibinfo {year} {2017})}\BibitemShut {NoStop}%
\bibitem [{\citenamefont {Du~Hamel~de
  Milly}(2017)}]{du_hamel_de_milly_manipulation_2017}%
  \BibitemOpen
  \bibfield  {author} {\bibinfo {author} {\bibfnamefont {X.}~\bibnamefont
  {Du~Hamel~de Milly}},\ }\emph {\bibinfo {title} {Manipulation of the mutual
  synchronisation in a pair of spin torque nano oscillators}},\ \href
  {http://www.theses.fr/2017SACLS442} {\bibinfo {type} {{PhD} thesis}},\
  \bibinfo  {school} {Paris Saclay} (\bibinfo {year} {2017})\BibitemShut
  {NoStop}%
\bibitem [{\citenamefont {Kolmogorov}(1954)}]{kolmogorov_conservation_1954}%
  \BibitemOpen
  \bibfield  {author} {\bibinfo {author} {\bibfnamefont {A.~N.}\ \bibnamefont
  {Kolmogorov}},\ }\href@noop {} {\bibfield  {journal} {\bibinfo  {journal}
  {Dokl. Akad. Nauk SSSR}\ }\textbf {\bibinfo {volume} {98}},\ \bibinfo {pages}
  {527} (\bibinfo {year} {1954})}\BibitemShut {NoStop}%
\bibitem [{\citenamefont {Arnold}(1963)}]{arnold_proof_1963}%
  \BibitemOpen
  \bibfield  {author} {\bibinfo {author} {\bibfnamefont {V.~I.}\ \bibnamefont
  {Arnold}},\ }\href@noop {} {\bibfield  {journal} {\bibinfo  {journal}
  {Uspekhi Mat. Nauk}\ }\textbf {\bibinfo {volume} {18}},\ \bibinfo {pages}
  {13} (\bibinfo {year} {1963})}\BibitemShut {NoStop}%
\bibitem [{\citenamefont {Moser}(1962)}]{moser_invariant_1962}%
  \BibitemOpen
  \bibfield  {author} {\bibinfo {author} {\bibfnamefont {J.}~\bibnamefont
  {Moser}},\ }\href@noop {} {\bibfield  {journal} {\bibinfo  {journal} {Nachr.
  Akad. Wiss. G{\"o}ttingen Math.-Phys. Kl. II}\ }\textbf {\bibinfo {volume}
  {1962}},\ \bibinfo {pages} {1} (\bibinfo {year} {1962})}\BibitemShut
  {NoStop}%
\bibitem [{\citenamefont {Louis}\ \emph {et~al.}(2018)\citenamefont {Louis},
  \citenamefont {Sulymenko}, \citenamefont {Tiberkevich}, \citenamefont {Li},
  \citenamefont {Aloi}, \citenamefont {Prokopenko}, \citenamefont {Krivorotov},
  \citenamefont {Bankowski}, \citenamefont {Meitzler},\ and\ \citenamefont
  {Slavin}}]{louis_ultra-fast_2018}%
  \BibitemOpen
  \bibfield  {author} {\bibinfo {author} {\bibfnamefont {S.}~\bibnamefont
  {Louis}}, \bibinfo {author} {\bibfnamefont {O.}~\bibnamefont {Sulymenko}},
  \bibinfo {author} {\bibfnamefont {V.}~\bibnamefont {Tiberkevich}}, \bibinfo
  {author} {\bibfnamefont {J.}~\bibnamefont {Li}}, \bibinfo {author}
  {\bibfnamefont {D.}~\bibnamefont {Aloi}}, \bibinfo {author} {\bibfnamefont
  {O.}~\bibnamefont {Prokopenko}}, \bibinfo {author} {\bibfnamefont
  {I.}~\bibnamefont {Krivorotov}}, \bibinfo {author} {\bibfnamefont
  {E.}~\bibnamefont {Bankowski}}, \bibinfo {author} {\bibfnamefont
  {T.}~\bibnamefont {Meitzler}}, \ and\ \bibinfo {author} {\bibfnamefont
  {A.}~\bibnamefont {Slavin}},\ }\href {\doibase 10.1063/1.5044435} {\bibfield
  {journal} {\bibinfo  {journal} {Applied Physics Letters}\ }\textbf {\bibinfo
  {volume} {113}},\ \bibinfo {pages} {112401} (\bibinfo {year}
  {2018})}\BibitemShut {NoStop}%
\end{thebibliography}%

\end{document}